\documentclass[preprints,article,accept,moreauthors]{mdpi}

\firstpage{1} 
\pubvolume{xx}
\issuenum{1}
\articlenumber{5}
\pubyear{2019}
\copyrightyear{2019}
\history{Received: date; Accepted: date; Published: date}

\usepackage{amsmath,amssymb}
\usepackage{physics}

\Title{Static and dynamic properties of a few spin $1/2$ interacting fermions trapped in a harmonic potential}

\Author{Abel Rojo-Francàs, Artur Polls and Bruno Juli\'a-D\'iaz}

\AuthorNames{Artur Polls, Bruno Julia-Diaz, Abel Rojo-Francàs}

\address{%
\quad Departament de Física Quàntica i Astrofísica, Facultat de Física, and 
Institut de Ciències del Cosmos (ICCUB), 
Universitat de Barcelona, E–08028 Barcelona, Spain}

\abstract{We provide a detailed study of the properties of a few interacting spin $1/2$ 
fermions trapped in a one-dimensional harmonic oscillator potential. The interaction 
is assumed to be well represented by a contact delta potential. Numerical results 
obtained by means of direct diagonalization techniques are combined with analytical 
expressions for both the non-interacting and strongly interacting regime. The $N=2$ case 
is used to benchmark our numerical techniques with the known exact solution of the problem. 
After a detailed description of the numerical methods, in a tutorial-like manner, we 
present the static properties of the system for $N=2, 3, 4$ and 5 particles, e.g. 
low-energy spectrum, one-body density matrix, ground-state densities. Then, we consider 
dynamical properties of the system exploring first the excitation of the breathing mode, 
using the dynamical structure function and corresponding sum-rules, and then a sudden 
quench of the interaction strength. }

\begin{document}


\tableofcontents
\clearpage
\section{Introduction}

The theoretical study of one-dimensional systems has always attracted a lot of attention. 
Very early in the development of quantum mechanics it was recognized that reducing the 
dimensionality of the systems enhances the quantum effects and gives origin to a plethora 
of interesting phenomena, see for instance Ref.~\cite{giamarchi} and references therein. 
After the pioneering work of Tonks, who investigated the equation 
of state of hard-rods, hard disks and hard-spheres in one, two and three 
dimensions, respectively~\cite{hard}, Girardeau established the map between impenetrable 
bosons and free fermions in one dimension~\cite{absolute.value}. However, it took a long 
time until the experimental realization of these systems became a reality~\cite{weis1,bloch1,weis2}.  
This experimental breakthrough took place in the context of the ultracold gases that 
have developed a frenetic experimental and theoretical activity during the last years 
achieving unbelievable possibilities to control the geometry, the interactions and the 
number of particles for many types of setups~\cite{bloch2,dalfovo,giorgini,sanpera}. 
More recently, the groundbreaking experiments of Jochim's group in Heildelberg, have 
opened new theoretical challenges  to study one-dimensional fermionic systems. They 
have been able to precisely control the number of atoms and the strength of 
the interactions~\cite{selim1}. These experiments have provided deep insight in 
the expected phenomena of fermionization of two distinguishable fermions~\cite{selim2}, 
i.e. with different third spin component. In addition, they have been able to change 
the number of particles, going from few to many particles, and being able to study the 
appearance of correlations in the many-body wave function. In fact, by adding particles 
to the system one by one, it was shown how the many-body correlated system is 
built~\cite{selim3}. These experiments used $^6$Li atoms and took advantage of the 
Feshbach resonance mechanism to modulate the interactomic 
interaction~\cite{feshbach.reson.ultracold}. Consequently, many theoretical problems 
have been addressed and few- and many-body techniques for ab-initio descriptions 
used in other fields have been specially tailored for these many-body one-dimensional 
systems. The impurity problem, the presence of pairing phenomena in few-fermion 
systems~\cite{selim4}, the behaviour of  two fermions in a double-well 
potential~\cite{selim5}, the realization of an antiferromagnetic spin chain of few 
cold atoms~\cite{selim6},  all belong  to the long list of theoretical challenges that  
the experimental advances are offering to theoreticians. They are looking at those 
systems as a very versatile laboratory where they apply  ab-initio techniques, without the 
complexity of the inter-particle interactions that appear in other fields, as for 
instance in nuclear physics~\cite{nuclei,nuclei2}.
Reciprocally, the  technological control is such that ultracold atomic systems can 
also be used as quantum simulators to solve very intricate 
condensed-matter problems~\cite{bloch2}. Although we will not pay attention 
to them in this paper it is worth to mention that trapping and control of the 
interactions have been extended also to mixtures of different nature and statistics: 
mixtures of bosons, and bosons and fermions~\cite{review_2019,boson.fermion,boson.boson,blume1,Schmelcher3,Mistakidis3}. 

In this work, we consider a system of a few spin-$1/2$ fermions trapped 
in a one-dimensional harmonic trap. The interaction between the fermions is 
modeled by a contact potential. In the literature there are some studies under 
similar conditions~\cite{harshman,review_2019,sun_fermions,1d_harmonically_sun,Sowinski_paper1,Zinner2, Zinner3,Sowinski3,Schmelcher2}. We 
pay attention both to the ground state and energy spectrum of the system 
and also to the dynamical properties of few-fermion systems depending on the 
interaction strength and the number of particles~\cite{dyn.struc.funct.,breathing_mode_bosons}.

The structure of this work is the following. In Section~\ref{chap1}, we present the 
theoretical description of the fermionic system, composed by a fixed number of atoms with 
spin up and down. In particular, we write down the full Hamiltonian both in first 
and second quantized form. Some limiting scenarios are considered, e.g. both zero 
and strong interaction. In Section~\ref{chap2}, we describe the numerical tools used 
to study the ground state and low-excited states of the fermionic mixture. The main 
method employed is direct diagonalization of the Hamiltonian matrix in a truncated 
Hilbert space. In Section~\ref{chap3}, we study the ground state of the system for 
different number of fermions and polarizations. In Section~\ref{chap4} we turn 
our attention to the lower part of the spectrum, computing the spectra for the same 
few-fermion systems considered. In Section~\ref{chap5}, we compute the response of the 
system to two different external perturbations. Finally in Section~\ref{chap6} we provide 
the main conclusions of our work.
 
\clearpage
\section{Theoretical approach}\label{chap1}

In this section we introduce the theoretical tools to describe a system composed of 
a few fermions trapped in a one dimensional harmonic oscillator potential. The goal is 
to provide a self-contained explanation of the methods employed, detailing the numerical 
subtleties encountered. 

\subsection{The Hamiltonian of the system}\label{sec1.1}

The Hamiltonian of the system can be split in two parts, the non-interacting single-particle 
part, which describes particles trapped in a harmonic oscillator potential, and the two-body 
interaction part, which describes the interactions between the fermions. The 
Hamiltonian, in first quantized form, can be written as~\cite{Sowinski_paper1,breathing_mode_bosons}
\begin{equation}\label{hamilton_0}
    H= H_{\mathrm{ho}}+\sum_{i<j}^N V_{{\mathrm{int}}}(x_i-x_j) \,,
\end{equation}
where $H_{\mathrm{ho}}= \sum_i h_{\mathrm{ho}}(i)$ is the sum of single-particle 
harmonic oscillator hamiltonians. The  Hamiltonian, $h_{\mathrm{ho}}$, is a 
one-body operator, which  includes the kinetic- and the harmonic-potential energy,
\begin{equation}\label{Hh.o.(i.s.units)}
    h_{\mathrm{ho}}=-\frac{\hbar^2}{2m}\frac{\partial^2}{\partial x^2}+\frac{1}{2}m\omega ^2 x^2\,,
\end{equation}
where $m$ is the particle mass and $\omega$ is the harmonic oscillator frequency. 
The eigenvalues of this Hamiltonian are well known,
\begin{equation}\label{h.o.energy}
    e_n=\left(n+\frac{1}{2}\right)\hbar\omega\;\;\; n=0,1,2,... \,,
\end{equation}
and the corresponding eigenfunctions are written as, 
\begin{equation}\label{h.o.w.f.(i.s.units)}
    \Phi_n(x)=\frac{1}{\sqrt{2^nn!}}\left(\frac{m\omega}{\pi\hbar}\right)^{1/4}
    e^{-\frac{m\omega x^2}{2\hbar}}H_n\left(\sqrt{\frac{m\omega}{\hbar}}x\right)\,,
\end{equation}
where $H_n(x)$ is the $n$-th Hermite polynomial. These single-particle wave 
functions will be used to build the many-body basis. All these wave functions 
have a similar structure: a Gaussian function multiplied by a Hermite 
polynomial with a normalization factor, that depends on $n$.

We consider identical spin $1/2$ fermions, i.e., with two possible spin 
states: $\ket{1/2,m}$, $m=1/2,-1/2$, where $m$ is the spin projection on the 
$z$-axis. Sometimes, these two spin states are denoted by: $\ket{\uparrow}$ 
and $\ket{\downarrow}$, respectively. The fermions are assumed to interact 
via a contact spin-independent delta potential~\cite{scat.lenght.}. Therefore 
two fermions interact only if they are at the same position. However, the 
many-body wave function of a system of identical fermions should be antisymmetric, 
preventing two fermions with the same spin from being at the same position. 
Therefore, fermions with the same spin projection do not interact, and 
thus the contact interaction is only acting between fermions with different 
spin projection.

In any case, the total wave function including the spin degree of freedom 
should be antisymmetric. This requirement allows the system to have two 
particles with different spin projection in the same position.

The interaction term of Eq.~(\ref{hamilton_0}) is a contact interaction, which 
is expressed as~\cite{scat.lenght.}
\begin{equation}
    V_{{\mathrm{int}}}(x_i-x_j)=g\delta(x_i-x_j) \,,
\end{equation}
where $g$ characterizes the strength of the interaction and $\delta$ is a 
Dirac delta function.

Taking into account the identity~\cite{dyn.struc.funct.}
\begin{equation}
    \sum_{i=1}^N x_i^2=NX^2+\frac{1}{N}\sum_{i<j}(x_i-x_j)^2\,,
\end{equation}
where $X=\frac{1}{N}\sum_i x_i$ is the center of mass coordinate, the 
Hamiltonian can be split in two pieces,
\begin{equation}
    H=H_{\mathrm{CM}}+H_{\mathrm{r}}\,,
\end{equation}
where $H_{CM}$ in harmonic oscillator units reads:
\begin{equation}\label{cm_hamiltonian}
    H_{CM}=-\frac{1}{2N}\frac{d^2}{dX^2}+\frac{N}{2}X^2\,,
\end{equation}
and governs the center-of-mass motion. $H_r$ affects only the relative 
coordinates and is translationally invariant.

One way to create a one-dimensional trap is using a cigar-shaped 
trapping potential, with the transverse trap frequency $\omega_\perp$ 
(in the radial direction) much larger than the trap frequency $\omega$ 
in the axial direction. In this situation, the coupling constant $g$ is 
related to the one-dimensional scattering length ($a_{\mathrm{1d}}$) 
as~\cite{exp.correlated_gas,scat.lenght.},
\begin{equation}
    g = -\frac{\hbar^2}{ma_{\mathrm{1d}}}\,.
\end{equation}

From the relation between the one-dimensional scattering length and the 
three-dimensional scattering length~\cite{dyn.magnet.}, we can write the 
coupling constant as
\begin{equation}
    g=\frac{2\hbar^2 a_{\mathrm{3d}}}
    {ma_\perp^2}\frac{1}{1-|\xi(1/2)|a_{\mathrm{3d}}/(\sqrt{2}a_\perp)}\,,
\end{equation}
with $a_\perp=\sqrt{\hbar/m\omega_\perp}$ and $\xi$ is the Riemann zeta function.

In a trap under these conditions, the system can be treated as a one-dimensional 
system. As the trapping in the transverse dimension is very strong, all particles 
occupy the lowest state of the transverse harmonic oscillator and the physics 
takes place in the longitudinal direction~\cite{dyn.magnet.}.

For convenience we use the harmonic oscillator units, in which the 
energy is measured in $\hbar \omega$ units, the length in units of 
$\sqrt{\hbar/m\omega}$ and the coupling constant $g$ is expressed in 
units of $\sqrt{\omega \hbar^3/m}$. From this moment on, all magnitudes will 
be expressed in these units.

\subsection{Non-interacting and infinite interaction limits}\label{sec1.2}

In general, it is not possible to solve analytically the Schr\"odinger equation 
for an arbitrary value of the interaction strength. However, there are two limits, 
i.e. the non-interacting and the infinite interaction cases, in which one has 
explicit analytical solutions~\cite{Exact.sol.,Zinner1}.

For these two limits, one can easily determine the energy and the density 
profile of the ground state of the system, and explicitly write the wave 
function.

\subsubsection{Non-interacting case: Fermi gas}
\label{subsec1.2.1}

In the non-interacting case, the system is in a state called Fermi gas. In 
these conditions, the behavior of the system is given by the single-particle 
states of the harmonic oscillator with the restriction of the Pauli principle 
that does not allow to have two fermions with the same spin projection in 
the same harmonic oscillator state. More specifically, in the ground state 
of a system with $N_d$ particles of spin down and $N_u$ particles of spin 
up, the particles occupy the lowest single-particle energy states, 
$N_d$ and  $N_u$, respectively. Therefore, the energy of the ground state 
is
\begin{equation}\label{E_Fermigas}
    E_{\mathrm{gs}}(g=0)=\sum_{n=0}^{N_d-1}\left(n+\frac{1}{2}\right)
    +\sum_{n=0}^{N_u-1}\left(n+\frac{1}{2}\right)=\frac{N_d^2+N_u^2}{2}\,.
\end{equation}
Actually the wave function of the system is given by a Slater determinant built 
with the lowest $N_u$ and $N_d$ single-particle wave functions with spin up 
and down, respectively.
The total third component of the spin of this wave function is $M=\frac{N_u-N_d}{2}$ 
and the total spin is the minimum $S$ compatible with $M$, $S=M$~\cite{min.spin}.
For a non-polarized system, $N_u=N_d$, then $M=0$ and $S=0$.

The density profile associated to this wave function is the sum of the 
probabilities of finding a particle in each occupied state in the position $x$,
\begin{equation}\label{density_Fermi}
    \rho(x)=\sum_{n=0}^{N_d-1}\abs{\Phi_n(x)}^2+\sum_{n=0}^{N_u-1}\abs{\Phi_n(x)}^2\,,
\end{equation}
where $\Phi_n(x)$ are the 1D harmonic oscillator wave functions, Eq.~(\ref{h.o.w.f.(i.s.units)}). 
The density profile is normalized to the number of particles.

\subsubsection{Infinite interaction case}
\label{subsec1.2.2}

In the infinite interaction limit, the system experiences a phenomenon reminiscent of the fermionization process in Bose systems~\cite{hard,absolute.value}. 
The ground state of the Hamiltonian in the infinite interaction limit for a fully- polarized state is a Slater determinant of the $N$ first harmonic oscillator states, $\Psi_A$. For a non fully-polarized case the ground state many-fold is degenerate, with degeneration $D=N!/(N_u!N_d!)$~\cite{Sowinski_paper1} and total third spin component $M=(N_u-N_d)/2$. The states can be written as~\cite{Zinner1},
\begin{equation}\label{Tonkswavefunc}
    \Psi=\sum_{k=1}^{D}a_k\theta(x_{P_k(1)},...,x_{P_k(N)})\Psi_A(x_1,...,x_N)\,,
\end{equation} 
where $\theta(x_1,...,x_N)=1$ when $x_1<x_2<...<x_N$ and zero otherwise, 
$P_k$ are the $D$ permutations of the $N$ components and the $a_k$ are the coefficients for each permutation. In this expression we have fixed the first $N_u$ particles to be spin projection up and the remaining $N_d$ particles to have spin projection down. Their 
energy depends only on the single-particle states present 
in $\Psi_A$, 
\begin{equation}\label{E_Tonksgas}
    E_{\mathrm{gs}}(g\rightarrow \infty)
    =\sum_{n=0}^{N-1}\left(n+\frac{1}{2}\right)=\frac{(N_d+N_u)^2}{2}\,.
\end{equation}

The density profile associated to these wave functions is the sum of the
probabilities of finding a particle in each occupied state in the position $x$~\cite{Decamp1},
\begin{equation}\label{density_Tonks}
    \rho(x)=\sum_{n=0}^{N-1} \abs{\Phi_n(x)}^2\,,
\end{equation}
where $\Phi_n(x)$ are the 1D harmonic-oscillator wave functions, 
Eq.~(\ref{h.o.w.f.(i.s.units)}).

\subsection{Second quantization}\label{sec1.3}

When working with several particles, it is useful to use the second-quantization 
formalism. One of the main ingredients of this formalism is the many-body Fock 
space. In order to define the Fock space, we need to have a single-particle basis, 
such as the harmonic oscillator states. The basis of the Fock space is constructed 
by indicating how many particles occupy each single-particle state. In 
addition, the antisymmetry rules are implemented by the anticommutation rules 
fulfilled by the creation and annihilation operators~\cite{second.quant.}.

A particular state is the vacuum state denoted by $\ket{0}$. This vector represents 
a state without particles, i.e., all its occupation numbers are zero. This state 
has norm equal to 1.

\subsubsection{Creation and annihilation operators}\label{subsec1.3.1}

The main tool in second quantization are the creation $(a_i^{\dagger})$ and annihilation 
$(a_i)$ operators, which act on the Fock space. Any observable, which is represented 
by an operator, can be expressed in terms of them~\cite{second.quant.}.
The action of a creation operator on the vacuum state, denoted as $a^\dagger _i\ket{0}$, 
creates a particle in the state $i$. In general, acting with a creation operator on 
a Fock space vector, creates a particle in the $i$-th state, whenever possible. 
Or in other words, it increases the occupation number of this single-particle 
state by one. On the other hand, the action of an annihilation operator ($a_i$) on 
a state destroys a particle in the $i$-th state, whenever possible. In other words, 
it decreases the occupation number of the state by one.

Also, a Fock space vector with $N$ particles can be created by the action of $N$ 
creation operators on the vacuum space, as follows,
\begin{equation}
    \ket{n_1,n_2,n_3,...,n_m}=
    (a^\dagger_1)^{n_1}(a^\dagger_2)^{n_2}(a^\dagger_3)^{n_3}...(a^\dagger_m)^{n_m}\ket{0}\,
\end{equation}
with $N=\sum_i n_i$.

In the case of fermions, to satisfy the Pauli principle, i.e., the antisymmetry of the 
wave function, the creation and annihilation operators should fulfill the following 
anti-commutation relations
\begin{equation}\label{anticomut}
\begin{split}
    &\left\{ a^\dagger_i,a_j\right\}=\delta_{ij} \\
    &\left\{ a^\dagger_i,a^\dagger_j\right\}=\left\{ a_i,a_j\right\}=0\,.
\end{split}
\end{equation}

Therefore the occupation number of the Fock basis can take the values $n_i=0$ 
or $n_i=1$. Finally, the action of the creation and annihilation operators can be 
written as
\begin{equation}
    \begin{split}
&a_i^\dagger \ket{n_1, n_2, ... n_{i-1}, n_i, ...}=
(-1)^{N_i} (1-n_i)\ket{n_1, n_2, ... n_{i-1}, (1-n_i), ...}\\
        & a_i \ket{n_1, n_2, ... n_{i-1}, n_i, ...}=
        (-1)^{N_i} n_i \ket{n_1, n_2, ... n_{i-1}, (1-n_i), ...}\,,
    \end{split}
\end{equation}
where $N_i$ is the number of occupied states with index lower than $i$,
\begin{equation}
    N_i=\sum_{k=1}^{i-1} n_k \,.
\end{equation}
The phase $(-1)^{N_i}$ is due to the anti-commutation relations. Notice that the factor 
in front of the state becomes zero if one tries to put two particles in the same 
single-particle state or one tries to annihilate a particle in a single-particle 
state which is not occupied.

\subsubsection{Fock space}\label{subsec1.3.2}

The harmonic oscillator eigenfunctions are used to build the many-body basis of 
the Fock space. In addition, the spin degrees of freedom are also incorporated 
to the single particle wave functions which are defined as: 
$\varphi_{n,m}(x,s)=\Phi_n(x)\chi_{m}(s)$, where $\Phi_n(x)$ is the harmonic 
oscillator wave function of the level $n$ and $\chi_m$ is the spin wave function, 
where $m=\uparrow(\downarrow)$ is the spin projection.

With these single-particle states, we build the many-body basis of the Fock space:
\begin{equation}
\ket{n_{0\downarrow},n_{0\uparrow},n_{1\downarrow},n_{1\uparrow},...}\,,
\end{equation}
where $n_{i,m}$ indicates the number of particles in the $\varphi_{i,m}$ single-particle state.

\subsubsection{Operators in second quantization}\label{subsec1.3.3}

In second quantization, operators can be expressed in terms of creation 
and annihilation operators. A general one-body operator $\hat{\mathcal{O}}$ 
is expressed as
\begin{equation}\label{1bop.sec.q.}
    \hat{\mathcal{O}}=\sum_{i,j} \bra{i}\hat{O}\ket{j} a_i^\dagger a_j \,,
\end{equation}
where $\ket{i}$ and $\ket{j}$ are single-particle states, 
$\bra{i}\hat{O}\ket{j}=\int\varphi_i^{*}(x)O(x)\varphi_j(x)dx$ is the one-body 
matrix element and $\varphi_i(x)$ are the single-particle wave functions used 
to build the Fock space. Notice that the sub-index $i$ stands for all the quantum 
numbers necessary to specify the single-particle state.

A general two-body operator $\hat{\mathcal{V}}$ is expressed as
\begin{equation}
    \hat{\mathcal{V}}=\frac{1}{2}\sum_{i,j,k,l} \bra{i,j}\hat{V}
    \ket{k,l}a_i^\dagger a_j^\dagger a_l a_k=\frac{1}{2}\sum_{i,j,k,l} v_{ij,kl} 
    a_i^\dagger a_j^\dagger a_l a_k \,,
\end{equation}
where $v_{ij,kl}$ is the two-body matrix element defined as
\begin{equation}
    v_{ij,kl}=\int \varphi_i^*(x_1)\varphi_j^*(x_2)\mathcal{V}(x_1,x_2)\varphi_k(x_1)\varphi_l(x_2) dx_1dx_2\,.
\end{equation}

\subsubsection{The Hamiltonian in second quantization}\label{subsec1.3.4}

As already mentioned, the Hamiltonian describing the harmonic oscillator is a 
one-body operator. On the other hand, the interaction term Eq.~(\ref{hamilton_0}) 
is a two-body operator. In our case, choosing  the  harmonic-oscillator eigenfunctions 
as the single-particle basis, the Hamiltonian part corresponding to the harmonic 
oscillator is diagonal, with eigenvalues $\epsilon_i=i+1/2$. Therefore in second 
quantization it reads, 
\begin{equation}
    \hat{h}_{\mathrm{ho}}=\sum_i \epsilon_i a^\dagger_i a_i =\sum_i \epsilon_i \hat{n}_i \,,
\end{equation}
where $\hat{n}_i=a_i^\dagger a_i$ is the number operator associated to the single-particle 
state $\ket{i}$.

For the interaction term, the two-body matrix elements are expressed as 
\begin{equation}\label{vijkl}
\begin{split}
    v_{ij,kl}=&\int dx_1dx_2\varphi^{*}_{i,{m}_i}(x_1,s_1)\varphi^{*}_{j,{m}_j}(x_2,s_2)
    g\delta(x_1-x_2)\varphi_{k,{m}_k}(x_1,s_k)\varphi_{l,{m}_l}(x_2,s_l)\\
    =&g\int dx \Phi_i^*(x)\Phi_j^*(x)\Phi_k(x)\Phi_l(x)\bra{\chi_{m_i}
    \chi_{m_j}}\ket{\chi_{m_k}\chi_{m_l}}\\
    =&g\delta_{{m}_i,{m}_k}\delta_{{m}_j,{m}_l}\int dx\Phi_i^*(x)\Phi_j^*(x)\Phi_k(x)\Phi_l(x) \,.
\end{split}
\end{equation}
Notice that the interaction does not affect the spin of the particles, and 
we have used the orthogonality of the spin functions: 
$\bra{\chi_{m_i}\chi_{m_j}}\ket{\chi_{m_k}\chi_{m_l}}=\delta_{{m}_i,{m}_k}\delta_{{m}_j,{m}_l}$. 
The calculation of the integral can be found in Section~\ref{integ.calc.}. Notice that 
the indices labeling the single-particle states run over the harmonic oscillator wave 
functions and the spin projections. Finally, the Hamiltonian reads
\begin{equation}
H=\sum_i \hat{n}_i \epsilon_i + \frac{1}{2}\sum_{ijkl}v_{ij,kl}a^\dagger _i a^\dagger_j a_l a_k\,.
\end{equation}

\clearpage
\section{Numerical methods}\label{chap2}

In this chapter we describe the method named direct diagonalization to find an approximate solution of the 
many-body Schr\"odinger equation. The technical aspects on how to construct the matrix 
Hamiltonian, to choose a suitable many-body basis of the Fock space, and how to calculate 
the two-body matrix elements are discussed in detail. Finally, we use the two-particle 
system, that has an analytical solution~\cite{Two.atoms}, as a benchmark of our numerical 
procedure.

\subsection{Direct  diagonalization }\label{sec2.1}

In order to study the ground state and the low-excited states of a system, one 
has to solve the many-body Schr\"odinger equation. With this objective, we use a 
direct diagonalization in a truncated basis space. This method needs to built the Hamiltonian matrix in 
an appropriate subspace and diagonalize it. In order to obtain accurate results we 
need to use a large basis, generating large matrices. To diagonalize these large 
matrices, we use a Lanczos method implemented by the ARPACK package~\cite{ARPACK}. 
This method allows us to diagonalize large matrices, and obtain the lowest eigenvalues 
with high accuracy. Hence, the direct  diagonalization method allows one to 
obtain the ground state and the low-energy spectrum and the corresponding states. Other strategies 
can also provide the lowest part of the spectrum, such as the MCTDH~\cite{cederbaum1}, used in Ref.~\cite{Schmelcher2} for few-fermion systems. In contrast, methods such as Monte Carlo~\cite{Braun1}, or DMRG~\cite{Bellotti1}, are usually used to compute ground state properties.

We consider systems with 2 to 5 fermions. These particles have spin 1/2, therefore 
for $N$ particles we can have $N+1$ possible total spin projections. As the Hamiltonian 
commutes with the total third spin component, the Hamiltonian is built in boxes with 
well defined $M$. Thus, it is possible to treat each total spin projection 
independently~\cite{Sowinski_paper1}. The simplest case is when all fermions have spin 
up, then $M=N/2$, and the wave function factorizes in an antisymmetric spatial wave 
function and a symmetric spin wave function $\chi (S=N/2, M= N/2)$ for $N$ particles. 
The antisymmetric spatial wave function is a Slater determinant built with the lowest 
single-particle states. The antisymmetry of the wave function prevents two particles 
from being in the same position and the particles in this wave function do not feel 
the contact interaction. Therefore, its energy is independent of the interaction strength.

The other trivial cases are the negative $M$ values because due to the spin symmetry 
of the Hamiltonian, the properties of the states do not depend on the sign of the 
spin projection. The diagonalization of a Hamiltonian box with a given $M$ would 
provide the same eigenvalues that the box with total spin projection $-M$. Therefore we 
concentrate in the study of cases with $M \ge 0$. 

\subsubsection{Basis truncation}\label{sec2.2}

To obtain the exact results, when using a diagonalization method, we would need to use 
a complete basis. However this is not possible because of the infinite dimension of 
the Hilbert space. Therefore we are forced to diagonalize the Hamiltonian in a finite 
subspace. Usually, this procedure does not provide the exact eigenvalue. However, the 
lowest-energy obtained by diagonalization is still an upper-bound to the ground-state 
energy. Note this is in contrast to exact diagonalization problems in small and discrete 
spaces, modelling finite optical lattices~\cite{traditional.trunc.}, where the method provides 
the exact values within machine precision. 

A common way to construct a finite many-body basis is considering a finite 
number of single-particle states, usually, those with the lowest 
energy~\cite{traditional.trunc.}. In our case, we diagonalize the Hamiltonian in a 
subspace with well-defined total third spin component. We start considering the 
lowest $N_M$ single-particle states and take into account an energy constraint in 
the construction of the many-body basis: the energy of the many-body basis states, 
which is given by the sum of the single-particle energies, should be smaller or 
equal than a fixed energy $E_{\mathrm{max}}$, this procedure is described 
in~\cite{basis.truncation,basis.truncation2}.

This maximum energy depends on the number of particles and spin configuration. 
Notice that we consider always $M>0$ therefore, $N_u>N_d$. The maximum energy, 
$E_{\mathrm{max}}$, corresponds to the energy of a non-interacting many-body 
state built as follows: one spin-up particle in the maximum single-particle energy 
state and the remaining ($N_u-1$) spin-up particles in the lowest $N_u-1$ 
single-particle states. On the other hand the $N_d$ spin-down particles are located 
in the lowest $N_d$ single-particle states. Therefore, the maximum energy considered 
in the construction of the many-body basis is
\begin{equation}
    E_{\mathrm{max}}=\sum_{n=0}^{N_d-1}\left(n+\frac{1}{2}\right)
    +\sum_{n=0}^{N_u-2}\left(n+\frac{1}{2}\right)+\left(\left(N_M-1\right)+\frac{1}{2}\right)
    =\frac{\left(N_d\right)^2+\left(N_u-1\right)^2}{2}+\left(N_M-\frac{1}{2}\right)\,,
\end{equation}
where $N_M$ is the number of single-particle states used.
\begin{table}[t]
\centering
\begin{tabular}{|l|c|c|c|}
\hline
\multicolumn{1}{|c|}{\multirow{2}{*}{}} & \multirow{2}{*}{\begin{tabular}[c]{@{}c@{}}number of\\ single-particle states\end{tabular}} & \multicolumn{2}{c|}{number of many-body basis
states} \\ \cline{3-4} 
\multicolumn{1}{|c|}{} &  & with energy restriction & without restriction \\ \hline
2 particles, M=0 & 100 & 5,050 & 10,000 \\ \hline
3 particles, M=1/2 & 50 & 10,725 & 61,250 \\ \hline
4 particles, M=0 & 40 & 30,800 & 608,400 \\ \hline
4 particles, M=1 & 40 & 19,530 & 395,200 \\ \hline
5 particles, M=1/2 & 30 & 22,923 & 1,766,100 \\ \hline
5 particles, M=3/2 & 30 & 11,349 & 822,150 \\ \hline
\end{tabular}
\caption{Number of single-particle states used in the construction of the many-body 
basis states in the second column. The number of many-body basis states, with and 
without energy restriction, are shown in the third and fourth columns, respectively.}
\label{tab:number_of_mode}
\end{table}

One can see in Table~\ref{tab:number_of_mode} that the dimension of the many-body 
basis is strongly reduced if the energy restriction discussed above is taken into 
account. Nevertheless, the reduction of the size of the space considered does not 
affect the quality of the results in the low-energy regime of the spectrum that 
we are exploring in this paper. The dimension of the many-body basis 
in the case without energy restriction is
\begin{equation}
    \frac{(N_M)!}{(N_M-N_u)!N_u!}\frac{(N_M)!}{(N_M-N_d)!N_d!}\,,
\end{equation}
which grows very rapidly.

In the table we have omitted the configurations with a maximum 
value of $M$, which correspond to trivial cases, with only one configuration built 
as the product of a symmetric spin function, with all spins parallel and an 
antisymmetric spatial function built with the lowest single-particle energy wave 
functions.  As it has been said previously, these configurations  do not feel the 
effects of a contact  interaction.

\subsection{The two-body matrix elements of the interaction}\label{integ.calc.}

The evaluation of the two-body interaction on the many-body Fock basis, requires 
the calculation of two-body interaction matrix elements, Eq.~(\ref{vijkl}). These 
matrix elements contain an integral with four wave functions, 
\begin{equation}
\begin{split}
    v_{ij,kl}=&g\delta_{m_i,m_k}\delta_{m_j,m_l}\int dx \Phi_i^{*}(x)\Phi_j^{*}(x)\Phi_k(x)\Phi_l(x)\\
    =&g\delta_{m_i,m_k}\delta_{m_j,m_l}I_{ijkl} \,.
    \end{split}
\end{equation}
The integral, that we label as $I_{abcd}$ can be solved analytically, because the 
wave functions are the harmonic oscillator wave functions Eq.~(\ref{h.o.w.f.(i.s.units)}),
\begin{equation}
\begin{split}
    I_{abcd} & =\int_{-\infty}^{\infty}\Phi_a(x)\Phi_b(x)\Phi_c(x)\Phi_d(x)dx \\
    & =\frac{1}{\pi\sqrt{2^{a+b+c+d}a!b!c!d!}}\int_{\infty}^\infty e^{-2x^2}H_a(x)H_b(x)H_c(x)H_d(x)dx \,.
\end{split}
\end{equation}
Let us consider, 
\begin{equation}
    I^{'}_{abcd}=\int_{-\infty}^\infty e^{-2x^2} H_a(x)H_b(x)H_c(x)H_d(x)dx\,.
\end{equation}
We notice that this integral has to be zero if ($a+b+c+d$) is an odd number. This is because 
the $\Phi_i$ wave functions have well-defined parity. Therefore if the multiplication of 
the four functions is odd, the integral value is zero. This is why we only calculate 
the integrals with ($a+b+c+d$) even.

Using the properties~\cite{Herm.integ.}
\begin{equation}\label{integral3hp}
\begin{split}
    &\int_{-\infty}^\infty e^{-2x^2}H_a(x)H_b(x)H_c(x)=\\
    &=\frac{2^{(a+b+c-1)/2}}{\pi}\Gamma\left(\frac{a+b-c+1}{2}\right)
    \Gamma\left(\frac{a-b+c+1}{2}\right)\Gamma\left(\frac{-a+b+c+1}{2}\right)\,,
\end{split}
\end{equation}
and
\begin{equation}\label{hprelation}
    H_m(x)H_n(x)=2^nn!\sum_{r=0}^n\frac{m!}{(n-r)!(m-n+r)!}\frac{H_{m-n+2r}(x)}{2^r r!} , n\leq m \,.
\end{equation}

We can express the integral $I_{abcd}^{'}$ as
\begin{equation}
\begin{split}
    I^{'}_{abcd}  =&2^d d! \sum_{r=0}^d \frac{c!}{(d-r)!(c-d+r)!}\frac{1}{2^r r!}\frac{2^{(a+b+c-d+2r-1)/2}}{\pi} \\
    & \times \Gamma\left(\frac{a+b-c+d+1}{2}-r\right)\Gamma\left(\frac{a-b+c-d+1}{2}+r\right)\\
    &\times \Gamma\left(\frac{-a+b+c-d+1}{2}+r\right)\,.
\end{split}
\end{equation}
Notice that the arguments of the gamma functions ($\Gamma$) are half-integers. Because 
($a+b+c+d$) is even, then when changing any sign, the result will remain even, and 
the gamma functions read,
\begin{equation}\label{gammavalues}
\left. \begin{split}
    &\Gamma\left(n+\frac{1}{2}\right)=\sqrt{\pi}\frac{(2n)!}{4^n n!} \\
    &\Gamma\left(\frac{1}{2}-n\right)=\sqrt{\pi}\frac{(-4)^n n!}{(2n)!}
\end{split}\right\}, n\geq 0 \,.
\end{equation}
Finally, the integral $I_{abcd}$ can be expressed as
\begin{equation}\label{Iabcd}
\begin{split}
    I_{abcd}=& \frac{1}{\pi^2\sqrt{2}}\sqrt{\frac{c!d!}{a!b!}}
    \sum_{r=0}^d\frac{1}{r!(d-r)!(c-d+r)!}\Gamma\left(\frac{a+b-c+d+1}{2}-r\right)\\
    &\times\Gamma\left(\frac{a-b+c-d+1}{2}+r\right)\Gamma\left(\frac{-a+b+c-d+1}{2}+r\right) \,.
\end{split}
\end{equation}

The presence of factorials in the expressions of the harmonic oscillator 
wave functions can cause fake overflows in the calculation. One possible solution 
to this problem is by taking logarithms of the expressions to be evaluated.
As shown before, a suitable way to write the integrals entering in the two-body 
matrix elements is
\begin{equation}
\begin{split}
    I_{abcd}=&\sum_{r=0}^d \frac{1}{\pi^2\sqrt{2}}\sqrt{\frac{c!d!}{a!b!}}\frac{1}{r!(d-r)!(c-d+r)!}\Gamma\left(\frac{a+b-c+d+1}{2}-r\right)\\
    &\times\Gamma\left(\frac{a-b+c-d+1}{2}+r\right)\Gamma\left(\frac{-a+b+c-d+1}{2}+r\right) \,.
\end{split}
\end{equation}
This expression can be written as $I=\sum_r^d f(a,b,c,d,r)$, where $f$ is a well-behaved 
function. Furthermore, we can write this result as
\begin{equation}
    I_{abcd}=\sum_{r=0}^d \exp{\log{[f(a,b,c,d,r)]}}\,,
\end{equation}
where $f$ must be positive.\\ \\
The logarithm $\log{(f)}$ is given by 
\begin{equation}
\begin{split}
    \log{(f)}=&-\frac{1}{2}\log{(2)}-2\log{(\pi)}\\ &+\frac{1}{2}\left(\log{(c!)}+\log{(d!)}
    -\log{(a!)}-\log{(b!)}\right)\\
    &-\log{\left((d-r)!\right)}-\log{\left((c-d+r)!\right)}-\log{(r!)}\\
    &+\log{\left(\Gamma\left(\frac{a+b-c+d+1}{2}-r\right)\right)}\\
    &+\log{\left(\Gamma\left(\frac{a-b+c-d+1}{2}+r\right)\right)}\\
    &+\log{\left(\Gamma\left(\frac{-a+b+c-d+1}{2}+r\right)\right)}\,.
\end{split}
\end{equation}
Notice that due to the presence of the gamma functions, the previous expression 
could require the evaluation of a logarithm of a negative quantity. For this 
reason, we calculate instead $\log{(\abs{\Gamma(n)})}$, and compute the associated 
phase separately. This phase is introduced in the final expression after the 
exponentiation. Thus, the final expression of the integral taking into account 
the possible negative values of gamma functions reads
\begin{equation}
    I_{abcd}=\sum_{r=0}^d p\exp{\log{[f(a,b,c,d,r)]}}\,,
\end{equation}
where $p$ is the phase, given by $p=\prod_{n=1}^3 p_n$, with $p_n$ being the phase 
generated by each gamma function.

\begin{figure}[t]
    \centering
    \includegraphics[width=\textwidth]{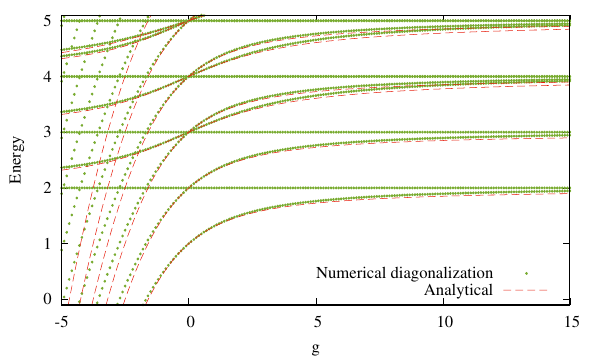}
    \caption{Energy spectrum of the two-particle system as a function of the 
    interaction strength. The calculations are performed using 100 single-particle 
    modes (green dots). The analytical spectrum (dashed red line) is obtained 
    with Eq.~(\ref{eq.bush}) for the energy of the relative system and adding after 
    the possible energies of the center of mass. Energy given in harmonic oscillator units. The energies obtained via Eq.~(\ref{eq.bush}) correspond to even-parity states, the corresponding odd-parity states do not depend on the interaction, and result in horizontal lines in close agreement with the numerical ones. }
    \label{fig:2partcomp}
\end{figure}
The logarithm of the gamma functions and the associated phases are calculated as
\begin{equation}
\left. \begin{split}
    & \log{\left(\Gamma\left(n+\frac{1}{2}\right)\right)}=\frac{1}{2}\log{(\pi)}+\log{\left((2n)!\right)}-2n\log{(2)}-\log{\left((n)!\right)}\;\;\;,\;\;\;p=1\\
    &\log{\left(\left|\Gamma\left(\frac{1}{2}-n\right)\right|\right)}=\frac{1}{2}\log{(\pi)}+2n\log{(2)}+\log{\left((n)!\right)}-\log{\left((2n)!\right)}\;\;\;,\;\;\;p=(-1)^{n}
\end{split}\right\}, n\geq 0\,.
\end{equation}
The logarithm of a factorial is calculated as:
\begin{equation}
    \log{(N!)}=\sum_{n=1}^N \log{(n)}\,.
\end{equation}

\subsection{A benchmark for the two-particle case}\label{sec2.5}

In general, it is not possible to solve exactly the energy spectrum. 
In our case, to determine the energy spectrum we have to diagonalize 
the Hamiltonian in a large Hilbert space by using sophisticated numerical 
techniques. However the case of two particles has been analytically 
solved in the literature~\cite{Two.atoms}.

\subsubsection{Theoretical spectrum for two particles}\label{subsec2.5.1}

For two fermions with opposite spin, $M=0$, the energies of the relative 
motion are obtained by solving the transcendental equation~\cite{Two.atoms}
\begin{equation}\label{eq.bush}
    \frac{\Gamma(-E_r/2+3/4)}{\Gamma(-E_r/2+1/4)}=-\frac{g}{2^{3/2}}\,,
\end{equation}
where $\Gamma$ are gamma functions, $E_r$ is the energy of the relative 
system and $g$ is the interaction strength.

In addition, the center-of-mass motion is governed by an harmonic oscillator 
Hamiltonian Eq.~(\ref{cm_hamiltonian}), and its energy is given by 
Eq.~(\ref{h.o.energy}). Then, the energy of a two-body state is the sum of 
its relative and center-of-mass energies. Therefore, each relative state has 
its corresponding center-of-mass excitations.

These analytical results are used as a test for our numerical calculations 
and allow us to critically analyze the dependence of the numerical results on 
the size of the Fock subspace used in the diagonalization. As $g$ increases one 
needs a larger subspace. 

\subsubsection{Comparison of analytical and numerical results}\label{subsec2.5.2}

In Fig.~\ref{fig:2partcomp} we report as a function of $g$ the low-energy part of 
the two-particle energy spectrum calculated by diagonalization in a truncated basis space (green dots) 
and the spectrum obtained by solving Eq.~(\ref{eq.bush}). Notice that this equation 
provides the energies of the relative motion to which we add the possible energies 
of the center of mass: $E_{CM}$. For the ground state, $E_{CM}=1/2$. On the contrary, 
the diagonalization provides the total energies of the system. The diagonalization 
has been performed using 100 single-particle modes that translates when the 
energy restriction is taken into account into a dimension of 5050 of the matrix 
to be diagonalized.

\begin{figure}[t]
    \centering
    \includegraphics[width=\textwidth]{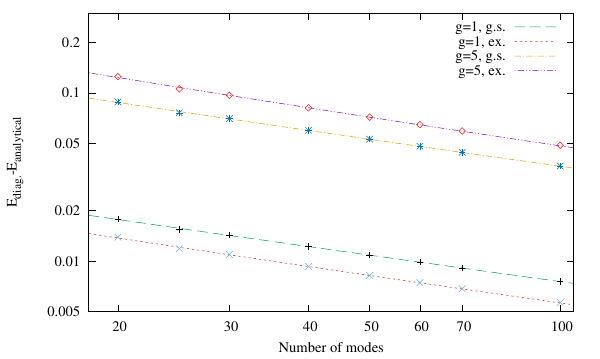}
    \caption{Differences between the numerical results and the exact ones as a function 
    of the number of  single particle modes, for two values of the interaction strength, 
    $g=1$ and $g=5$, for the ground (g.s) and first excited (ex.) state. The lines are fits of 
    the type $C_1/N_M^{1/2}+C_2/N_M$ to the calculated points. Both axis 
    are in a logarithmic scale.}
    \label{fig:error_fit}
\end{figure}
We explore both the attractive and the repulsive interaction regimes. In general, 
for small values of $|g|$, both attractive and repulsive, the agreement between 
both types of calculations is very good. Apparently, the quality of the results 
deteriorates faster for attractive interaction. The calculations discussed in 
this paper will consider mainly repulsive interactions which will be explored 
up to the interaction strength limit which is almost reached for the highest 
values of $g$ considered, $g=15$.

In the plot, we also include the horizontal lines which are energies of states not 
affected by the interaction. The first one corresponds to the state with $S=1$ and 
$M=1$ whose wave function can be decomposed as an antisymmetric wave function in 
coordinate space built with the single-particle states: $n=0,1$ times the triplet 
spin state with $M=1$. The energy of this state, $E=2$, does not depend on $g$. 
In fact, all the states described by horizontal lines in the figure, are eigenstates 
of the Hamiltonian with $M=1$. Notice also that in the limit $g \rightarrow \infty $ 
states with $M=0$ become degenerate with states with $M=1$.

To complete the study of the accuracy of our calculations we investigate the convergence 
of the energy of the ground state and first excited state of the two particle system 
as a function of the number of harmonic oscillator modes used ($N_M$). Note that for  
$N=2$, the energy constraint for the construction of the two-body basis is given by 
the maximum energy $E_{\mathrm{max}}= 1/2+ (N_M- 1/2) = N_M$.

To this end, in Fig.~\ref{fig:error_fit} we report the difference between the ground-state 
energy obtained by the diagonalization procedure and the analytical ones, for two values 
of the interaction strength. The figure also reports the difference of the first 
excited state. For $N=2$, it is possible to establish the dependence of the dimension 
of the Hilbert space on $N_M$, which is given by $N_M (N_M+1)/2$. As expected, the difference 
between the calculated and the exact value decreases with the number of modes. 
Following Ref.~\cite{chemest.}, we fit functions of the type $C_1/N_M^{1/2}+C_2/N_M$ 
with excellent results. In addition, for large values of $N_M$, the convergence of 
the energy goes as $1/N_M^{1/2}$. This dependence indicates a slow convergence of the 
numerical results by increasing the number of single-particle modes. As expected, 
the differences are larger for the larger strength, but in both cases the differences 
are very small. In all cases the difference is positive, indicating that the 
numerical results are upper bounds to the exact ones.

\clearpage
\section{Ground state properties}\label{chap3}

In this section we  discuss and analyze different observables characterizing the 
ground state of the system. This analysis is done for several numbers of particles 
and total spin projections, devoting special attention to the two-particle system, 
as it is the only one with analytical solution.

As the interaction is spin independent, the Hamiltonian for a given number of 
particles is calculated in a subspace with a well-defined total spin projection $M$. 
After diagonalization of the Hamiltonian, the lowest eigenvalue and its corresponding 
eigenvector is identified as the ground state of the system for this spin projection. 

The ground state is written in the Fock basis as
\begin{equation}
    \ket{\Psi}=\sum_n C_n \ket*{\psi_n}\,,
\end{equation}
where $\ket*{\psi_n}$ is a vector of the many-body basis of the Fock space, all 
with the same spin projection and constructed with the single particle eigenstates 
of the harmonic oscillator.

\subsection{Energy and virial theorem}\label{sec3.1}

\begin{figure}[t]
    \centering
    \includegraphics[width=\textwidth]{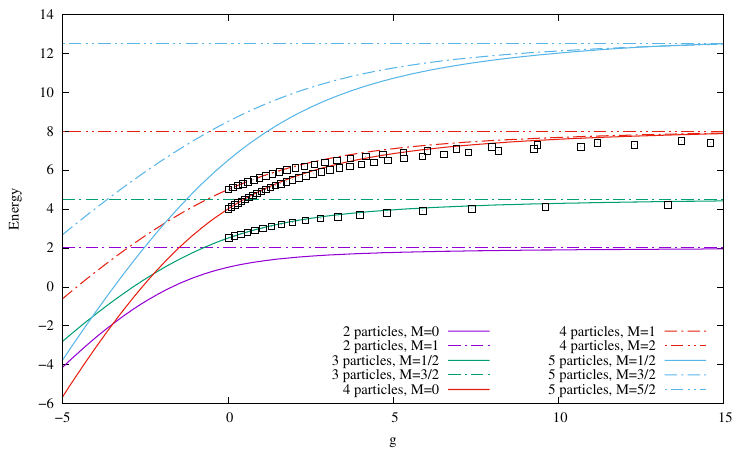}
    \caption{Ground-state energy as a function of the interaction strength for 
    different number of particles and spin configurations. The calculations have 
    been performed by using 100, 50, 40 and 30 single-particle modes to construct 
    the many-body basis for N=2,3,4 and 5 particles respectively. The black squares are the corresponding results reported in Ref.~\cite{blume2}. }
    \label{fig:gs.energy}
\end{figure}

The energy of the ground-state as a function of the interaction strength $g$ for 
several number of particles and spin configurations is shown in Fig.~\ref{fig:gs.energy}. 
The first thing to observe is that for each number of particles the states with 
maximum spin projection are not affected by the interaction. In fact, the wave function 
of these states can be factorized as a symmetric spin function, with all spins up 
and an antisymmetric spatial wave function. Due to the Pauli principle, this spatial 
wave function is built with the first $N$ single-particle harmonic oscillator 
states. As discussed in section 2 its energy is given by $E= N^2/2$. In all other 
cases, the ground-state energy grows when $g$ increases and tends 
to saturate when $g \rightarrow \infty$. More precisely, for the $N=2$, $M=0$, the 
energy evolves from $E=1$ at $g=0$, to $E=2$ in the limit $g \rightarrow \infty$. 
Later in this section, we will discuss the structure of both the spatial and the 
spin part of the wave function in both limits. For $N=3$, $M=1/2$ the energy goes 
from $E=5/2$, to the $E= 9/2$ for $g \rightarrow \infty$. In a similar way, for 
$N=4$, $M=0$ the energy goes from $E=4$ at $g=0$ to $E=8$ at $g \rightarrow \infty$ 
while for $N=4$, $M=1$ the energy goes from $E=5$ to $E=8$. Notice that in the 
limit $g\rightarrow \infty$ the energy depends only on the number of particles and 
it is the same for the different spin projections. The values of the energy at 
$g \rightarrow \infty$ were predicted by Eq.~(\ref{E_Tonksgas}), 
$E= (N_d+N_u)^2/2$ where $N_d$ and $N_u$ are the number of particles with spin up 
and down respectively. On the other hand, in the case of an attractive interaction, 
the binding energy increases as the interaction becomes more attractive. In this 
case the energies belonging to different number of 
particles cross in the figure, as the systems with more particles can accumulate 
more attraction.

In Fig.~\ref{fig:gs.energy} we also show the results for the cases $N=3$, $M=1/2$, $N=4$, $M=0$ and $N=4$, $M=1$ reported in Ref.~\cite{blume2} for the repulsive regime. For small interaction strengths, $g \lesssim 5$, both calculations agree fairly well, but for larger interactions we can appreciate that our calculations provide larger energy values due to the use of a truncated basis. This effect is more pronounced  in the four particles case, which is performed with less harmonic oscillator single-particle states.

After this general view of the dependence of the ground-state energy on the 
strength of the interaction for different systems, we are going to analyse in 
more detail the two-particle system with $M=0$, which energy as a function 
of $g$ is plotted in Fig.~\ref{fig:gs.energy}. In the non-interacting case, 
the wave function can be factorized
\begin{equation}
    \Psi_{g=0}(1,2)=\Phi_0(x_1)\Phi_0(x_2)\chi(S=0,M=0)\,,
\end{equation}
as a symmetric function in space with both particles in the same single-particle 
state and the two-body spin function corresponding to a singlet state which is 
antisymmetric: then the total wave function is antisymmetric and its energy is 
$E=1$. In the limit of a very strong interaction, the wave function can be factorized 
as~\cite{absolute.value}
\begin{equation}\label{infinite2part}
    \Psi_{g=\infty}(1,2)=\frac{1}{\sqrt{2}}|\Phi_0(x_1)\Phi_1(x_2)-\Phi_1(x_1)\Phi_0(x_2)|\chi(S=0,M=0)\,
\end{equation}
i.e., the absolute value of the determinant built with the ground and first 
excited single-particle states of the confining harmonic oscillator times the 
singlet two-particle spin function. In this case, the presence of the absolute value 
ensures that the spatial part of the wave function is symmetric. Notice also 
that this wave function does not allow two particles to be at the same 
position and therefore the two particles do not feel the interaction. The 
final energy of this wave function is the sum of the single-particle 
energies, $E=1/2+3/2=2$. However, even if the wave function of 
Eq.~(\ref{infinite2part}) looks rather simple, due to symmetry arguments, 
the diagonalization procedure requires a very large Fock space to 
asymptotically approach the solution.

It is also interesting to study the decomposition of the energy in different 
pieces: kinetic, harmonic-oscillator potential energy and interaction energy, 
which are constrained by the virial relation:
\begin{equation}\label{Virial}
    2\expval{T}-2\expval{V_{\mathrm{ho}}}+\expval{V_{\mathrm{int}}}=0\,.
\end{equation}
The fulfilment of this relation reinforces the accuracy of the calculations. The 
derivation of the virial relation can be found in the Appendix~\ref{apendix1}.
\begin{figure}[t]
    \centering
    \includegraphics[width=\textwidth]{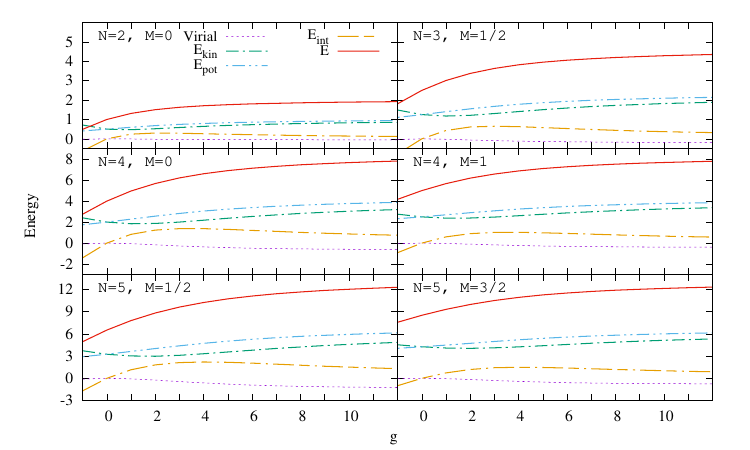}
    \caption{Different energy contributions to the total ground-state energy, as a function of the interaction strength for different number of particles and total third-spin component. The total energy and the fulfilment of the virial theorem is also shown (Eq.~(\ref{Virial})). The legend is common to all graphics.}
    \label{fig:virial_all}
\end{figure}

In Fig.~\ref{fig:virial_all} we report, as a function of the interaction strength, 
the different contributions to the ground-state energy, for different number 
of particles and total third-spin component. The total energy is also reported. 
In general, for all cases considered, the behavior of the different contributions 
is rather similar. For a repulsive interaction, the total energy increases. In the 
repulsive interaction range, both the kinetic energy and the oscillator-potential 
energy are similar. Actually, they are equal at $g=0$  and also tend to the same 
value in the limit of infinite interaction. The oscillator-potential energy is 
greater than the kinetic energy in this range. The interaction energy starts 
from zero in the non-interacting case, has a maximum, and goes to zero as the 
interaction strength tends to infinity.

Obviously, the virial theorem is trivially respected for the systems with maximum 
spin projection, as in these cases the interaction energy is zero and therefore 
the kinetic and harmonic potential energy coincide. In general the virial theorem 
is well fulfilled, although it deteriorates a little when the interaction strength 
is increased. The behavior of the different contributions 
to the total energy can be explained by taking into account the virial theorem and knowing 
the non-interacting and infinite-interacting limits. In the non-interacting case, 
the kinetic energy is equal to the oscillator potential energy. In the infinite 
interaction limit, the wave function is given by Eq.~(\ref{infinite2part}), that 
prevents two particles from having the same position, therefore, the interaction 
energy is zero. As a consequence, the kinetic and the harmonic oscillator energies 
are also equal and fulfill the virial relation. For intermediate interactions, 
the interaction energy is positive, and the kinetic energy is lower than the 
oscillator-potential energy in agreement with the virial theorem. The continuity 
of the interaction energy between zero and infinite interaction allows us to 
predict the existence of a maximum of the interaction energy as a function of 
$g$. On the other hand, when the interaction is attractive, the interaction energy 
takes negative values, increasing the binding energy as the interaction becomes 
more attractive. At the same time, the kinetic energy grows and the harmonic 
potential energy decreases, as a consequence of the decrease of the size of the 
system.

\subsection{One body density matrix}\label{sec3.2}

The one-body density matrix (OBDM) provides a non-trivial insight on the many-body 
structure of the system. The OBDM allows one to obtain the natural orbits, i.e. 
the eigenvectors of the OBDM, the momentum distribution and the density profile of 
the system. In second quantization, the matrix elements of the OBDM associated to 
$\ket{\Psi}$ are defined as,
\begin{equation}
    \rho_{ij}=\bra{\Psi}a_j^\dagger a_i \ket{\Psi}=
    \sum_{n,m}C_{n}^{*}C_{m}\bra*{\psi_n}a_j^\dagger a_i \ket*{\psi_m}\,.
\end{equation}

\subsubsection{Density profile}\label{subsec3.2.1}

The first observable to be calculated is the density profile of the ground 
state of the system, that provides information on how the particles are spatially 
distributed in the trap.

The density profile in terms of the matrix elements of the OBDM is given by, 
\begin{equation}
    \rho(x)=\sum_{i,j}\Phi_i^{*}(x)\rho_{ij}\Phi_j(x)\,,
\end{equation}
where $\Phi_n(x)$ are the single-particle wave functions used to construct 
the many-body basis. This expression corresponds to the diagonal elements of 
the one-body density matrix in spatial representation.
\begin{figure}[t]
    \centering
    \includegraphics[width=\textwidth]{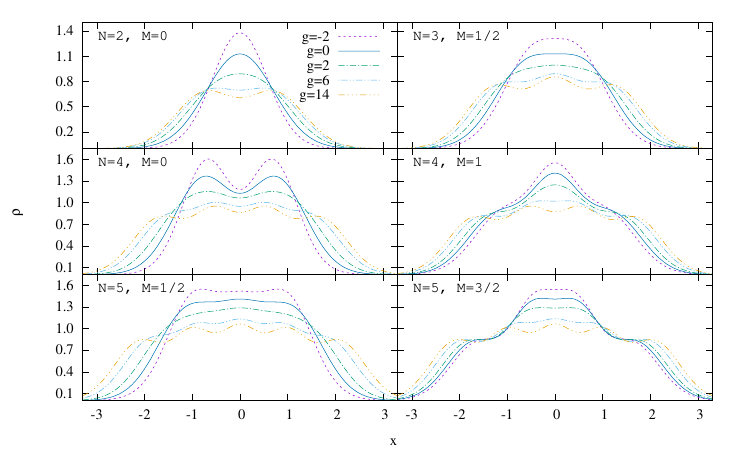}
    \caption{Density profiles of the ground state for different number of particles 
    and spin projections. The density profiles are shown for various values of the 
    interaction strength. The legend is common to all panels.}
    \label{fig:density_all}
\end{figure}

In Fig.~\ref{fig:density_all} we report the density profile of the ground state 
of the system for different number of particles and spin configurations and for 
several values of the interaction strength. As expected, the density profiles at 
$g=0$ for the different systems fully agree with the analytical expressions given in Eq.~(\ref{density_Fermi}). For large values of $g$, $g=14$ in the figure, the  
density profiles are very close to the ones predicted in Eq.~(\ref{density_Tonks}). 
The small differences are due to the finite value of $g$ and also to the finite 
size of the Hilbert space where we are diagonalizing the Hamiltonian. In any case, 
both for the energy and the spatial distribution of the particles, $g=14$ can be 
considered a strongly interacting regime. Notice, that for the same number of 
particles and different spin configurations the density profiles are rather 
different, when $g$ is small. However, for $g \rightarrow \infty$ the profile 
for a given number of particles does not depend on the spin projection.  In the 
strongly interacting limit, the density profile shows as many peaks as the number 
of particles, i.e. the particles try to be distributed equidistantly to minimize 
the repulsion. The size of the system also increases with the interaction strength. 
For attractive interactions, $g=-2$ is shown in the figure, the density profile 
is narrower and reaches higher values than in the repulsive case. 

\begin{figure}[t]
    \centering
    \includegraphics[width=\textwidth]{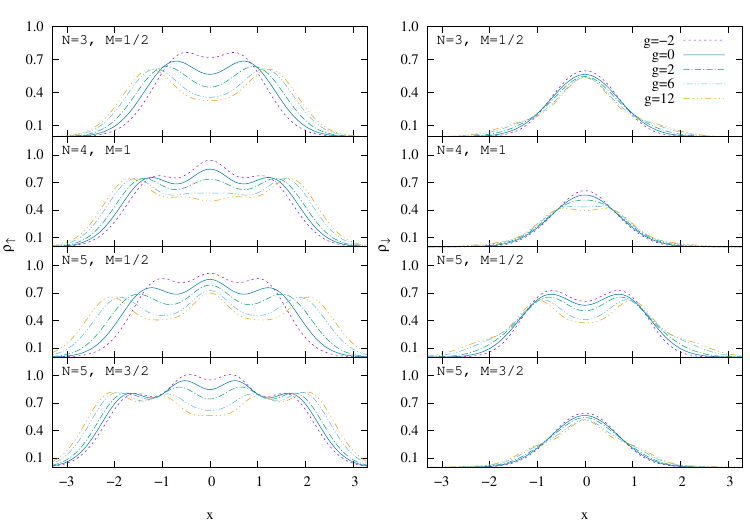}
    \caption{Densities of the spin up ($\rho_\uparrow$, left panels) and spin down ($\rho_\downarrow$, right panels) particles in the ground state. The density profiles are shown for various values of the interaction strength and for different number of particles and spin projection. The legend is common to all panels. The corresponding total densities are reported in Fig.~\ref{fig:density_all}.}
    \label{fig:density_spin}
\end{figure}

In Fig.~\ref{fig:density_spin} we report the densities of the spin up and spin down particles. The cases $N=2$, $M=0$ and $N=4$, $M=0$ are not reported, because these cases have the same particles with spin up and spin down, and for symmetry reasons, the densities of both types of particles are equal and these cases do not provide new information. For the remaining cases, we can see that in every density profile there are the same number of peaks as particles with the correspondent spin projection. As the interaction  increases, the peaks separate to mimimize the interaction energy.

\subsubsection{Natural orbits}\label{subsec3.2.2}

The diagonalization of the density matrix provides its eigenvectors, i.e. natural orbits, 
and its eigenvalues, which can be interpreted as the occupation number of the 
natural orbits. The sum of the eigenvalues of the natural orbits is normalized 
to the total number of particles.

In the non-interacting case, the ground state of an $N-$particle system, which 
corresponds to a Slater determinant built with $N$ single-particle harmonic oscillator 
functions with their spin projection, produces a OBDM with $N$-natural orbits with 
eigenvalue $1$ and the all others with eigenvalue zero. In these cases the natural 
orbits can be identified with the single-particle wave functions used in the 
construction of the Slater determinant. When turning on the interaction, one gets 
a set of $N$-natural orbits with eigenvalues smaller than $1$ and additional 
natural orbits with eigenvalues significantly smaller. The natural orbits are 
expressed as linear combinations of the harmonic oscillator single-particle 
basis and the Slater determinant built with the $N$ natural orbits with the 
highest eigenvalues define the wave function of this type with largest overlap 
with the ground state. However, this condition does not imply that the energy 
corresponding to this wave function is the minimum energy for 
a wave function built with a single Slater determinant. The fact that the OBDM has eigenvalues smaller 
than $1$ points out the existence of correlations beyond the mean-field 
in the ground-state of the system which translate into the impossibility to 
express the ground-state wave function as a single Slater determinant of single-particle wave functions.

The eigenvalues associated to the natural orbits can be interpreted as the 
occupation numbers of the single-particle states defined by the natural orbits. 
Notice also that a given natural orbit does not mix single particle states 
with different parity or spin projection.
\begin{figure}[t]
    \centering
    \includegraphics[width=\textwidth]{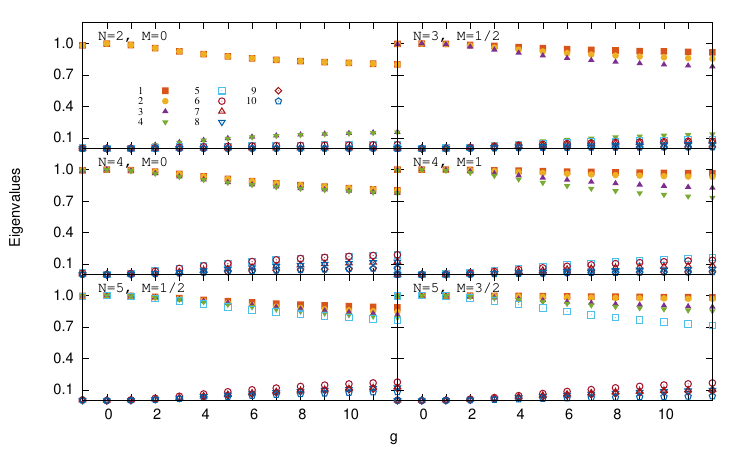}
    \caption{The largest ten eigenvalues of the OBDM of the ground state of systems 
    with different number of particles and spin configurations as a function of 
    the interaction strength.}
    \label{fig:obdm_eigenvalues}
\end{figure}

In Fig.~\ref{fig:obdm_eigenvalues} we  report the dependence of the ten largest  
eigenvalues of the one-body density matrix as a function of the interaction strength 
for the ground-state of systems with different number of particles and spin 
projections. As explained above, at $g=0$ we have $N$, where $N$ is the number 
of particles, eigenvalues equal to unity, and the rest equal to zero. When the 
interaction increases, the highest eigenvalues decrease indicating the presence 
of correlations in the system. Then as the eigenvalues are normalized 
to $N$, the smaller eigenvalues start to increase, indicating the impossibility 
to describe the wave function with only one Slater determinant. The fact that 
the eigenvalues of the OBDM  could be the same, does not necessarily imply 
that the natural orbits are the same. However, for $N=2$ and $M=0$ the 
eigenvalues appear doubly degenerate independently of the value of $g$. One 
corresponds to spin up and the other to spin down that have identical natural 
orbits and also the same spatial distribution.  The same is true for the $N=4$ and 
$M=0$ case. For $g=0$ we have four natural orbits,  two corresponding  to 
the single-particle state $n=0$, both with spin up and spin down and 
the other to $n=1$ also with both spin projections, all with eigenvalue 
equal to unity. Then when the interaction increases, the natural orbits mix 
higher excited harmonic oscillator states but do not mix the spin projection. 
The natural orbits are independent of the spin projection and the eigenvalues 
separate in two groups each one with degeneracy two. They have not only the 
same eigenvalue but also the same spatial structure. Instead, for $N=3$ and 
$M=1/2$ the natural orbits of each spin projection are different.

The reason behind the previous observations is that the OBDM is constructed in boxes 
of well defined third spin component,  
\begin{equation}
    \left(
\begin{split}
    &\rho_{\uparrow\uparrow}\;\; \rho_{\downarrow\uparrow}\\
    &\rho_{\uparrow\downarrow}\;\; \rho_{\downarrow\downarrow}
\end{split}
    \right)
    = \sum_{n,m}C_n^{*}C_m\bra*{\psi_n}\left(
\begin{split}
   &a^\dagger _{\uparrow}a_{\uparrow}\;\;&a^\dagger _{\downarrow} a_{\uparrow} \\
    &a^\dagger _{\downarrow}a_{\uparrow}\;\;&a^\dagger _{\downarrow} a_{\downarrow}
\end{split}
    \right)\ket*{\psi_m}\,.
\end{equation}
However, the crossed terms are zero and only the diagonal terms contribute. 
Therefore the matrix separates in two pieces, one with spin up and the other 
with spin down. If the system is symmetric in the third spin component, both 
subspaces are equal and the eigenvalues and spatial eigenfunctions of each 
subspace are also equal. 

\clearpage
\section{Low-energy excited states}\label{chap4}

To complete the study of the structure of the few-body fermion systems, in this 
section we discuss the low energy excitation spectrum which will be useful to 
understand the dynamics and the response of the system to an external perturbation.
In this section we present an algorithmic procedure to assign quantum labels to
  the different states, a more formal study of the symmetries in few-atom systems with
  examples for three to five atoms is given in Ref.~\cite{harshman}.

\subsection{Energy spectrum}\label{sec4.1}

\begin{figure}[t]
    \centering
    \includegraphics[width=\textwidth]{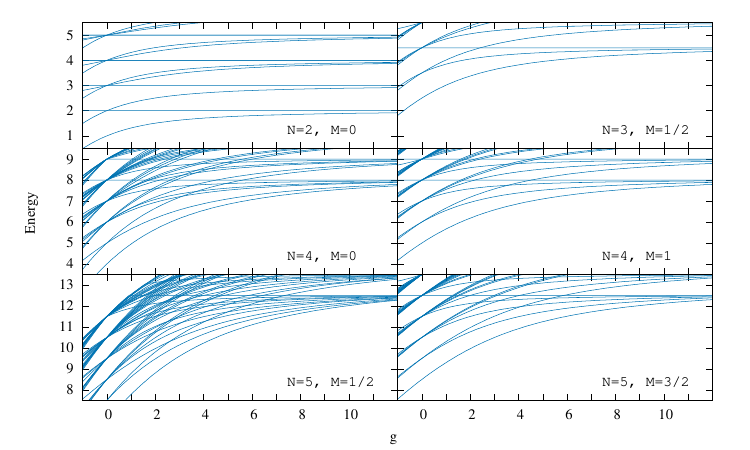}
    \caption{Low-energy spectrum for several number of particles and spin 
    configurations as a function of the interaction strength.  Notice that the excitation energies are measured with respect to 
    the ground-state energy of the non-interacting system.}
    \label{fig:spectrum}
\end{figure}

Once the Hamiltonian for a given number of particles ($N$) and total third spin 
component ($M$) has been diagonalized, we have access to the spectrum and the 
structure of the eigenstates. We pay particular attention to the dependence on the 
interaction strength of the low-energy part of the spectrum, which is shown in 
Fig.~\ref{fig:spectrum} for several number of particles and spin configurations.

The first observation is the existence of states which are not affected by 
the interaction and therefore appear as horizontal lines in the plots. These states 
correspond to antisymmetric wave functions in space and give zero probability of having 
two particles in the same position. Actually, these wave functions can be factorized 
in an antisymmetric wave function in space and a spin symmetric wave function. 
Later on, in Sect.~\ref{sec4.2} we discuss how to assign the total spin 
quantum number to the different eigenstates.

Another general observation is the existence of excitations with a constant 
energy shift independent of the interaction strength which is associated to excitations 
of the center of mass of the system. Notice that the interaction which is invariant 
under space translations does not affect the center-of-mass motion.

An interesting feature is that the energy of all states saturates when the 
interaction strength tends to infinity. Actually, some states merge to the same 
energy when $g\rightarrow \infty$ increasing the degeneracy of the energy 
levels in this limit. 

Once more, let us analyze more carefully the case $N=2$ and $M=0$. As explained in 
the previous section, the non-interacting ground state corresponds to 
\begin{equation}
    \Psi(1,2)=\Phi_0(x_1)\Phi_0(x_2)\chi(S=0,M=0)\,,
\end{equation}
with energy $E=1$. The two first excited levels have the same energy, $E=2$ at $g=0$, 
one corresponds to a state 
\begin{equation}
    \Psi(1,2)=\frac{1}{\sqrt{2}}(\Phi_0(x_1)\Phi_1(x_2)-\Phi_1(x_1)\Phi_0(x_2))\chi(S=1,M=0)\,,
\end{equation}
which is a product of an antisymmetric function in coordinate space and a symmetric one in spin. 
This state is not affected by the contact interaction and its energy is independent 
of $g$. The energy of this state merges with the ground state energy when 
$g\rightarrow \infty$. As the Hamiltonian commutes with the spin, this state 
is degenerate with states having the same spatial wave function and the spin 
functions corresponding to $S=1$ but $M=0,\pm 1$.

The other excited state at $g=0$ corresponds to
\begin{equation}
    \Psi(1,2)=\frac{1}{\sqrt{2}}(\Phi_0(x_1)\Phi_1(x_2)+\Phi_1(x_1)\Phi_0(x_2))\chi(S=0,M=0)\,,
\end{equation}
which is symmetric in spatial coordinates and antisymmetric in spin. It can be shown that 
this state corresponds to the first excitation of the center of mass times the intrinsic 
wave function of the ground state described above. In the ground state, the center of 
mass is always in the ground state of $H_{\mathrm{CM}}$. However, in the state we are 
considering the center of mass occupies the first excited state of $H_{\mathrm{CM}}$.  
This is true for all values of $g$ and the energy of this state can be written as 
$E=E_{g.s.}+1$ for all values of $g$. This state has degeneracy one. Actually, 
the wave function of this state can be written as the ground-state wave function 
at a given $g$ times the wave function of the first excited state of $H_{\mathrm{CM}}$

This type of arguments explain the full spectrum for $N=2$. It is worth reminding that 
our numerical procedure is constructed in the Fock space using single-particle wave 
functions, without decomposing the Hamiltonian in the center of mass and intrinsic 
parts. Therefore the proper localization of the center of mass excitations provides 
a test on the numerical accuracy of the calculations.

We continue our analysis by considering the case for $N=3$ and $M=1/2$. The ground 
state at $g=0$ has an energy  $E=1/2+1/2+3/2 = 5/2$, which corresponds to a state that 
populates the single-particle harmonic-oscillator states: 
$\{0 \uparrow, 0 \downarrow, 1 \uparrow\}$, that results in $M=1/2$. Then when $g$ 
increases, the state mixes  more complicated configurations,  the energy increases 
monotonously and tends to $E_{g \rightarrow \infty}= 1/2+3/2+5/2=9/2$. The spin of the 
ground state is $S=1/2$, which is the minimum compatible with the value of $M=1/2$ and 
its parity is negative all along the values of $g$. At  $g \rightarrow \infty $ the 
ground state merges with the state with $S=3/2, M=1/2$ which is a state built with a 
Slater determinant with the single-particle harmonic-oscillator states: $n=0,1,2$ 
times the symmetric wave function of three spins $S=3/2, M=1/2$. The particles in this 
state do not feel a contact interaction and the energy of this state is independent 
of the interaction strength. Notice also the existence of two states at $g=0$ with 
$M=1/2$ built with the single-particle harmonic-oscillator states 
$\{0 \uparrow, 0 \downarrow, 2 \uparrow\}$ and $\{0\uparrow, 1 \downarrow, 1 \uparrow\}$ 
which give origin at $g=0$ to two states with total energy $E=7/2$. One describes an 
excitation of the center of mass built on top of the ground state which evolves 
with $g$ keeping always the same energy shift respect to the ground state, and the 
other state merges with the ground state when $g \rightarrow \infty$.

Next, we discuss the case $N=4$ and $M=0$. Of course, the number of levels increases 
with the number of particles and the number of levels is larger when the total spin 
projection, $M$ is smaller. One immediately detects  states not affected by the 
interaction. In particular, the lowest one  that corresponds to a state with $S=2$, 
the maximum spin for $N=4$ and $M=0$. This state is described by  a wave function 
that is the product of an antisymmetric wave function given by a Slater determinant 
built with the single-particle harmonic-oscillator states: $n=0,1,2,3 $ and a 
symmetric spin function of four particles with $S=2$ and $M=0$. The total energy 
of this state, $E=1/2+3/2+5/2+7/2 = 8$, is given by the sum of the first four 
single-particle energies of the harmonic-oscillator trapping potential.

The ground state is a state with $S=0$, i.e. the minimum total spin compatible 
with the value of $M$. The energy of the state at $g=0$, $E=4$, corresponds to 
the occupation of the single particle levels: 
$\{0 \uparrow, 0 \downarrow, 1 \uparrow,1 \downarrow\}$. The  energy gets more 
repulsive as $g$  increases  and for $g \rightarrow \infty $ the energy merges 
with the energy $E=8$ of the previously  discussed state. The parity of the 
ground-state is positive. One can also identify the excitations of the 
center of mass  characterized by a constant shift of the energy respect to the 
ground-state or on  top of a given state. For instance, the first one corresponds 
to a center of mass excitation of one unit of energy of the Hamiltonian associated 
to the center of mass, $H_{\mathrm{CM}}$. There are other levels whose quantum 
numbers  can be identified and many of them merge together when $g \rightarrow \infty$.

Similar comments apply to the case $N=4, M=1$. One can also identify  states not 
affected by the interaction. In particular, the lowest one corresponds to  $S=2, M=1$, 
that has the same energy as the state $S=2, M=0$ of the previous panel. Actually, 
both states $S=2,M=1$ and $S=2,M=0$ can be built from the state $S=2,M=2$ constructed 
with an antisymmetric space wave function times  a symmetric spin wave function with 
all spins up. These states are obtained by applying the ladder operator $S_{-}$ to 
the state $S=2,M=2$. When the Hamiltonian commutes with $S_{-}$ this operation 
does not change the energy of the state. These arguments are discussed at 
length in Sect.~\ref{sec4.2}. In this $M-$box, the ground state starts with $E=5$ at 
$g=0$ which corresponds to the Slater determinant built with the single-particle 
states: $\{0 \uparrow, 0 \downarrow, 1 \uparrow, 2 \uparrow\}$ and when 
$g \rightarrow \infty$ merges with several states that tend to $E=1/2+3/2+5/2+7/2=8$.

Finally let us discuss very briefly the case $N=5, S=1/2$. In this case, there are  
many levels below the first level which is not sensitive to the interaction strength, 
with an energy $E=1/2+3/2+5/2+7/2+9/2=25/2$ and $S=5/2, M=1/2$. The ground state, 
has $S=1/2$, and positive  parity. At $g=0$ it has an energy $E=13/2$ and when 
$g \rightarrow \infty$ it merges with the state that is non-sensitive to the interaction 
discussed above, with an energy $E=25/2$.

For the case with $M=3/2$ there are considerably fewer levels. The ground state at 
$g=0$ has more energy than the ground state for $M=1/2$, basically due to 
the Pauli Principle. In fact, at $g=0$ the ground-state is built with the 
single particle states $ \{0 \uparrow, 0 \downarrow, 1 \uparrow, 2 \uparrow, 3 \uparrow \}$, 
which gives an energy $E=17/2$. However, when $g$ is increased the state merges 
with the state $S=5/2, M=3/2$ which is also degenerate with the sate 
$S=5/2, M=1/2$ discussed in the previous panel.

Whenever possible, we have compared our results with the ones provided 
in Ref.~\cite{Sowinski_paper1} obtaining a good agreement in all cases 
considered. In fact our results, which provide upper 
bounds, are in general slightly below the results of Ref.~\cite{Sowinski_paper1}.

Finally, let us point out that in the limit of large interaction several 
groups of states merge to the same energy giving rise to an increase of the 
energy degeneracy. In particular, for the ground state this degeneracy is
expressed as $D=\frac{N!}{N_u!N_d!}$. 

\subsection{Spin determination}\label{sec4.2}

As the Hamiltonian does not depend on the spin, it commutes with the spin 
operators and should be possible to assign a total spin to the eigenstates of the Hamiltonian. In particular, it commutes with the ladder lowering $S_-$ and raising $S_+$ spin 
operators. Therefore, if we have a state with a total spin $S$ and spin 
projection $M$ with energy $E$, due to the commutation relations of $S_{+(-)}$  
with the Hamiltonian, it turns out that by applying the operators $S_{+(-)}$ 
to this state one obtains a state with the total spin projection increased 
(reduced) by 1, i.e, $M \pm 1$, with the same energy $E$. Therefore, for each 
eigenstate with a well-defined spin, there will be $2 S+1$ states with the 
same energy. We can use this fact to determine the total spin of the states. 
The argument is as follows: Given $N$ fermions, we start considering the maximum 
possible spin projection, that will be the state with all spins pointing up, with 
$M=N/2$. In this case, the wave function factorizes in a spatial antisymmetric 
wave function times a symmetric spin wave function. This wave function has a well defined spin which 
coincides with the maximum value of the spin projection, 
$S=N/2$. Then if we apply $2S$ times to this state the operator $S_-$, we 
will get the remaining states belonging to this  spin multiplet, all having 
the same energy and all of them factorized as an antisymmetric wave function 
in space times a spin symmetric wave function. Obviously, these states are 
found in different $M$-box subspaces. In fact, if  we consider  the Hamiltonian 
in the box with $M = S-1$, among  the energies obtained in this box, we 
find the previous energy attributed to $S= N/2$, corresponding in this 
case to the state with $M=N/2-1$.

Besides, if the remaining states in this $M$-box are not degenerate, one can assure
that the total spin of each one will be
the minimum $S$ compatible with the value of $M$, i.e. $S=N/2-1$. 
One can have more than one state with this spin and one would 
need another quantum number to distinguish between them. Next, we take the 
box of the Hamiltonian associated to the value of $M$, $M=N/2-2$. Here we 
immediately identify the energies obtained for the previous spin values 
and again the rest of states have a spin that is the minimum compatible with 
the value of $M$. One can continue this process until the total spin projection 
has the positive lowest half-integer value for odd $N$ or zero value in 
case of even $N$.
\begin{figure}[t]
    \centering
    \includegraphics[width=\textwidth]{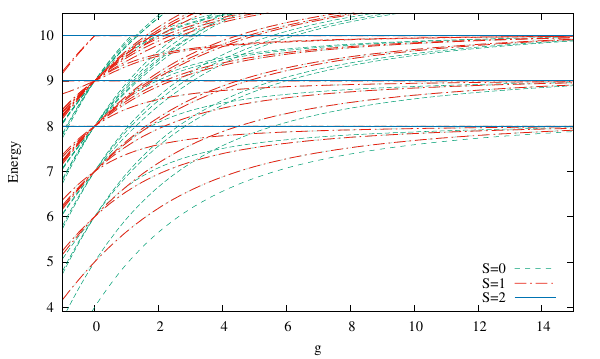}
    \caption{Low-energy spectrum of the fermionic system with four particles as a 
    function of the interaction strength. 
    The states are labelled according to their total spin, as explained in the text.}
    \label{fig:4particlesspin}
\end{figure}

To illustrate this procedure, we consider the lowest-energy states of the case 
$N=4$ shown in Fig.~\ref{fig:4particlesspin} as a function of the interaction 
strength. The maximum possible spin projection is $N/2=2$. The wave functions 
with $S=2$ and $M=2$ can be factorized in an antisymmetric spatial wave function times a spin wave function with all spins pinting up. In particular, the lowest energy for this states ($S=2$ and $M=2$), $E=8$, is the sum of the 
first four single-particle  harmonic oscillator energies, $E=1/2+3/2+5/2+7/2$.
In addition, this state is not affected by the interaction and therefore 
appears as the lowest horizontal line in figure. This energy
 level corresponds 
also to the energy of the different possible  $2S+1$ states of the $S=2$ 
multiplet. In the next step we consider the box with $M=1$. First we identify 
the previous energy and find out several non-degenerate eigenstates which can all be associated 
to $S=1$ (dot-dashed red lines in the figure). Notice that all of them have 
different energies, but $S=1$. For instance, at $g=4$ we find three $S=1$ levels 
with energies below $E=8$. In the next step we diagonalize the subspace with 
$M=0$ and identify  the $M=0$ states corresponding to $S=2$ and to the $S=1$ 
states detected before. The rest of levels have $S=0$. At $g=5$, we find  three 
$S=0$ levels below $E=8$, all with different energy, and non degenerate. Notice 
that one of them corresponds to a center of mass excitation and that merges 
with other levels at $E=9$   when $g\rightarrow \infty$. Using  these types of 
arguments one can assign the spin to the eigenstates of the Hamiltonian even if the 
total wave function could be complicated and most of the times non-factorizable 
in a spin and a space part.

\clearpage
\section{Dynamical excitation}\label{chap5}

In Sects.~\ref{chap3} and~\ref{chap4} we have concentrated on the static properties 
of the fermionic system. In particular, we have analyzed the ground state and 
also the main features of the low-energy spectrum. In the present section we turn 
our attention to the dynamics of the fermionic system in the harmonic trap. 
Simulating the dynamics of quantum many-body systems is important for 
current quantum technological applications where one could for instance 
design quantum systems to produce desired many-body states after a certain 
evolution~\cite{global_optimi.,quench.bos.,breath.bos.}. A second relevant 
aspect is that the dynamics of the system reflects in many ways its internal 
structure. In this sense, one can devise a dynamical evolution in order to unveil 
the spectral structure of the system~\cite{exciting_breathing_mode}. This is the 
main goal of this section. To this aim we consider two different perturbations 
to the ground state of the system, which are sensitive to the low-energy 
spectra presented before. The first is a sudden change in the trap 
frequency. This perturbation excites both center-of-mass and relative modes 
of the system and can be analyzed by computing the dynamic structure 
function. In the second one, we numerically obtain the time evolution of the 
system after a sudden quench of the interaction strength. In this case, 
the center-of-mass modes are not excited and we look for traces of the internal 
structure on the time evolution of the central density of the cloud.

\subsection{Sudden change in the trap frequency: breathing mode}\label{sec5.1}

A well-known way to study the internal structure of a quantum many-body 
system trapped in a harmonic oscillator is by exciting the so called breathing 
mode. In contrast with the Kohn (dipole) mode, where the cloud is initially 
displaced from the minimum of the harmonic potential~\cite{kohn.mode}, in our 
case we study the response of the system to a change in the trapping frequency.
For the two particle case there are studies as~\cite{Blume3,Ebert1}, and using different confining potential as~\cite{Mistakidis2}. 
The main tool we consider is the dynamic structure function associated 
to the mono-polar excitation operator
\begin{equation}\label{excitation_function}
    \hat{F}=\sum_j^N x_j^2\,.
\end{equation}
This perturbation preserves spin and parity, therefore, it only connects the 
ground state of the system with other excited states with the same total spin 
and parity.

This operator can be separated in two pieces: a center-of-mass and an intrinsic 
one, i.e. $\hat{F}=NX^2_{\mathrm{CM}}+\frac{1}{N}\sum_{j<i} (x_i-x_j)^2$. The 
center of mass of the system is described by an harmonic oscillator Hamiltonian, 
and the wave functions associated to this part are the harmonic oscillator wave 
functions. Note that this perturbation can excite the center of mass or the 
intrinsic part, but not both at the same time.

\subsubsection{Dynamic structure function}

The dynamic structure function, normalized to the number of particles, is 
defined as
\begin{equation}\label{structure_function_0}
    S_F(E)=\frac{1}{N}\sum_{i> 0} 
    \abs{\bra{\Psi_i}\hat{F}\ket{\Psi_0}}^2\delta \left(E-(E_i-E_0)\right)\,,
\end{equation}
where $\hat{F}$ is the excitation operator associated to the external perturbation, 
see Eq.~(\ref{excitation_function}), $\ket{\Psi_0}$ is 
the ground state of the unperturbed system and $\ket{\Psi_i}$ are its excited states.

In the second quantization formalism, the structure function reads,
\begin{equation}\label{structure_function_1}
    S_F(E)=\frac{1}{N}\sum_{i>0}\abs{\sum_{n,m}C_{n,0}C^{*}_{m,i}
    \sum_{k,l}\bra{k}x^2\ket{l}\bra*{\psi_m}a^\dagger_k a_l\ket*{\psi_n}}^2
    \delta\left(E-(E_i-E_0)\right)\,,
\end{equation}
where $\ket{k}$ and $\ket{l}$ are harmonic-oscillator single-particle states 
and $\ket*{\psi_n}$ are the Fock states of our basis. For the center of mass, 
the $\bra{i_{\mathrm{CM}}}X_{\mathrm{CM}}^2\ket{j_{\mathrm{CM}}}$ are different 
from zero only when $i_{\mathrm{CM}}=j_{\mathrm{CM}}$ or  
$i_{\mathrm{CM}}=j_{\mathrm{CM}}\pm 2$. This implies that the center of mass 
can be excited at most by two energy quanta.

In some limiting cases we know the analytic value of the dynamic structure 
function. These cases can be used to benchmark our 
numerical calculations~\cite{dyn.struc.funct.}. For the non-interacting case, 
we expect a single peak in the dynamic structure function with an energy 
$E-E_0=2$. The intensity of this peak depends on the number of particles 
and the total spin projection. Likewise, in the infinite interaction limit, 
we also expect a single peak in the dynamical structure function with 
energy $E-E_0=2$. In addition, a peak with energy $E-E_0=2$ is expected for 
any value of the interaction strength, due to the center-of-mass excitation, 
with a constant intensity as a function of the interaction strength. In 
the range of repulsive interaction, we expect more peaks associated to 
states with the same spin and parity as the ground state. These correspond 
to intrinsic excitations. 
\begin{figure}[t]
    \centering
    \includegraphics[width=\textwidth]{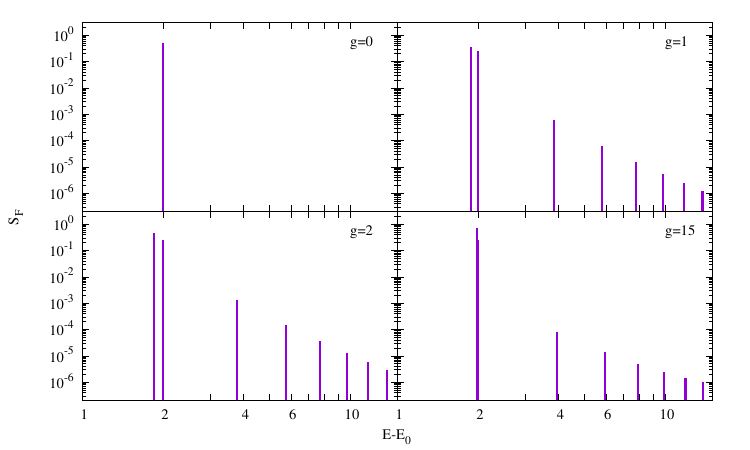}
    \caption{Dynamic structure function of a mono-polar excitation for the case 
    of two fermions. The total spin of the ground state and of all excited states 
    connected to it through the mono-polar excitation, are zero. The different 
    panels correspond to four different values of interaction strength. From the 
    non-interacting case, $g=0$ to a strongly interacting one $g=15$.}
    \label{fig:SF_breathing2p}
\end{figure}

In Fig.~\ref{fig:SF_breathing2p} we present the dynamic structure function 
for the two fermions, for the non-interacting case and for values of the 
interaction strength $g=1$, $g=2$, and $g=15$. We are considering the total 
spin projection $S=0$, i.e. the ground state and all the states excited by 
the monopolar operator have total $S=0$ and even parity. As expected, 
for the non-interacting case we only have one peak at excitation energy 
$E-E_0=2$. For any interaction strength there is a peak at $E-E_0=2$ with 
the excitation energy associated to the center-of-mass excitation with 
an intensity independent of the interaction strength. The rest of the 
peaks observed are associated to intrinsic excitations. In addition, for 
large interaction strengths, we recover only one peak at $E-E_0=2$, and the 
rest of peaks vanish. The intensity of the excitations decreases for the 
states with more energy, and there is a dominant peak near the energy 
of the center-of-mass excitation. In the non-interacting case, the single 
peak has two contributions: the center of mass, and the dominant intrinsic 
peak. When we turn on the interaction, the dominant intrinsic peak moves 
to a lower energy, but for large interaction, returns to the same energy 
than the center of mass excitation. This reentrant behaviour, reported for 
bosons in Ref.~\cite{reentrant.bos.}, is also reflected in the energy 
spectrum for the two particles case. For the remaining excitations, we 
can identify the correspondent states via the energy spectrum: all of 
them are excitations of the relative motion, with the same spin and 
parity than the ground state. Notice that the good location of the 
center-of-mass excitation is a good test on the numerical evaluation 
of $S_F(E)$, which in Eq.~(\ref{structure_function_1}) is expressed 
in terms of single-particle matrix elements and the Fock basis. It is 
also worth to mention that in this case $S_F(E)$ for two particles 
with $M=0$ is the same as if the particles were two bosons Ref.~\cite{dyn.struc.funct.}. 
In the case of $N=2$ with $M=0$, it has been possible to identify 
the center-of-mass and intrinsic excitations, because we know the 
analytic spectrum~\cite{Two.atoms}. This is not the case for the rest 
of configurations. The $N=4$ with $M=0$ configuration is a good example 
to explore the behaviour of the dynamic structure function of the 
breathing mode in a more complex case.
\begin{figure}[t]
    \centering
    \includegraphics[width=\textwidth]{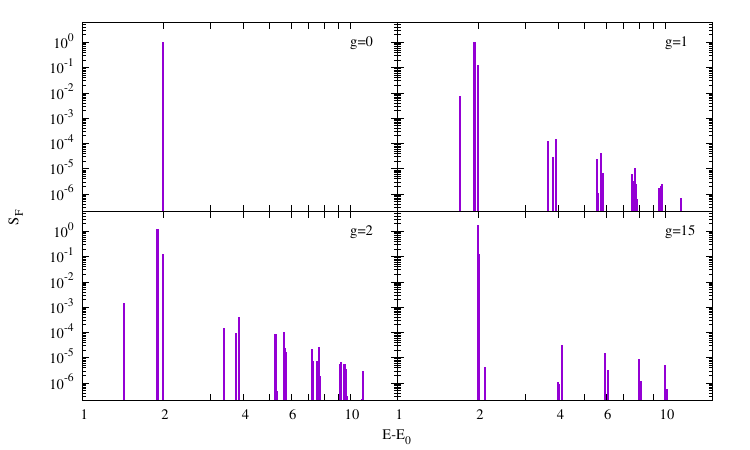}
    \caption{Dynamic structure function of a mono-polar excitation for the case 
    of four particles. The total spin of the ground state and of all excited states 
    connected to it through the mono-polar excitation, are zero. The different 
    panels correspond to four different values of interaction strength. From 
    the non-interacting case, $g=0$ to a strongly interacting one $g=15$.}
    \label{fig:SF_breathing4p}
\end{figure}

In Fig.~\ref{fig:SF_breathing4p} we report the dynamic structure function 
associated to the breathing mode for $N=4$ with $M=0$, for different values 
of the interaction strength, concretely, for the non-interacting case ($g=0$), 
for $g=1$, $g=2$ and $g=15$. As in the case of two particles, in the 
non-interacting case, there is only one peak with energy $E-E_0=2$. In the 
interacting case, there is a peak at energy $E-E_0=2$ with a constant 
intensity, which we identify with a center-of-mass excitation. All the 
peaks present can be associated to excitations of the ground state to states 
with the same spin and parity as the ground state. Using this information, 
we verify that the spin determinations in Sect.~\ref{sec4.2} are in 
agreement with the results obtained here. Note that not all the states that 
verify these conditions (have the same spin and parity as the ground state) 
are excited. This is because the state has to be an intrinsic excitation 
or a center-of-mass excitation. In addition, in this case we also observe 
a dominant peak with a reentrant behaviour. This is common to all the cases 
with different number of particles and total spin projection we have 
considered. Finally, we can see that in the limit of strong interaction, 
the intensity of all the peaks, except the dominant and the center of 
mass, decrease, tending to zero for the infinite interaction limit.

\subsubsection{Sum rules}

\begin{figure}[t]
    \centering
    \includegraphics[width=\textwidth]{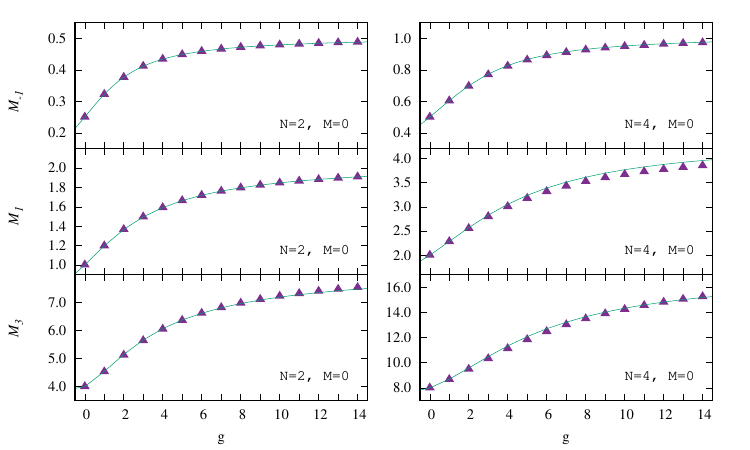}
    \caption{Values of the three energy momenta $M_{-1}$, $M_1$ and $M_3$ for the 
    cases of two and four particles as a function of the interaction strength. The 
    calculations done using the explicit value of the dynamic structure 
    function, Eq.~(\ref{energy_moment_0}) are represented by dots and the line 
    is the value computed using the sum rules, Eq.~(\ref{sum_rules_relation}). 
    For $M_{-1}$ the results from Eq.~(\ref{sum_rules_relation}) and Eq.~(\ref{mm1alt}) 
    are indistinguishable on the figure.
    \label{fig:sum_rules}}
\end{figure}

The energy momenta of the dynamic structure function are a useful tool to 
analyze the response of the system to an excitation. They are defined as
\begin{equation}\label{energy_moment_0}
    M_n=\int E^n S_F(E) dE\,,
\end{equation}
where $E$ is the excitation energy. They can be explicitly computed from the 
definition of $S_F(E)$, Eq.~(\ref{structure_function_1}), by using the second 
quantization formalism as:
\begin{equation}\label{energy_moment_1}
    M_n=\frac{1}{N}\sum_{i>0}\abs{\sum_{n,m}C_{n,0}C^{*}_{m,i}
    \sum_{k,l}\bra{k}x^2\ket{l}\bra*{\psi_m}a^\dagger_k a_l\ket*{\psi_n}}^2(E_i-E_0)^n\,.
\end{equation}
If the dynamic structure function is known, the energy momenta 
can be computed directly using Eq.~(\ref{energy_moment_0}). But in general, the 
dynamic structure function is difficult to calculate and is not exactly known. 
In these situations there are useful theorems that allow us to compute $M_n$ 
using only ground state properties~\cite{sum.rules}. These relations are named 
sum rules.

For the monopolar excitation operator, the lowest sum rules can be simply 
expressed as the following expectation values on the ground 
state~\cite{dyn.struc.funct.}
\begin{equation}\label{sum_rules_relation}
\begin{split}
    &M_{-1}=\left.-\frac{1}{2}\frac{1}{N}
    \frac{\partial^2E_0(\lambda)}{\partial \lambda^2}\right|_{\lambda=0}\\
    &M_{1}=\frac{4}{N}\expval{V_{\mathrm{ho}}}\\
    &M_{3}=\frac{4}{N}\left(\expval{T}+3\expval{V_{\mathrm{ho}}}\right)\,.
\end{split}
\end{equation}
Using perturbation theory, we have an alternative expression for $M_{-1}$,
\begin{equation}
    M_{-1}=\left.-\frac{1}{2}\frac{1}{N}
    \frac{\partial}{\partial \lambda}\bra{\bar{0}}\hat{F}\ket{\bar{0}}\right|_{\lambda=0}\,,
    \label{mm1alt}
\end{equation}
where $\ket{\bar{0}}$ and $E_0(\lambda)$ are the ground state and its corresponding 
energy of the perturbed Hamiltonian $\hat{H}^{'}=\hat{H}+\lambda\hat{F}$.

\begin{figure}[t]
    \centering
    \includegraphics[width=\textwidth]{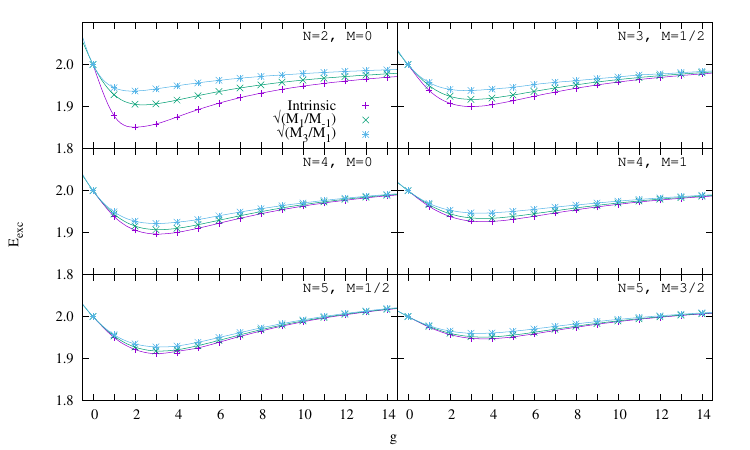}
    \caption{Ratios of the sum rules $\sqrt{M_{1}/M_{-1}}$ and $\sqrt{M_{3}/M_{1}}$ 
    as a function of the interaction strength for several number of particles and 
    spin configurations. The purple $+$ signs correspond to the excitation energy 
    of the main peak of the dynamic structure function.}
    \label{fig:reentrant}
\end{figure}
In Fig.~\ref{fig:sum_rules} we report the energy momenta $M_{-1}$, $M_1$ and $M_3$ 
as a function of the interaction strength. The sum rule $M_{-1}$ has been calculated 
only using the explicit values of the dynamic structure function. For $M_1$ and 
$M_3$, the two methods, direct integration of the dynamic structure function and 
sum rules, are in perfect agreement for the case of two particles. For $N=4$ the 
agreement between both methods is also very good. In all cases, the three momenta 
considered grow monotonically as the interaction strength is increased.

The sum rules can also be used to obtain approximate values of the excitation 
energies. This is a tool which is for instance of interest for quantum 
Monte Carlo methods, where the low-energy spectrum of the system cannot be 
computed, see for instance the bosonic case in~\cite{reentrant.bos.}. Indeed, 
if only one peak appears in the dynamic structure function, we can compute 
its excitation energy by $\sqrt{M_n/M_{n-2}}$~\cite{sum.rules}. In a more 
general case this ratio can at least be used as an estimation of the energy 
of the main peak.

In Fig.~\ref{fig:reentrant} we report the ratios $\sqrt{M_{1}/M_{-1}}$, 
$\sqrt{M_{3}/M_{1}}$ and excitation energy of the main contribution, which 
is associated to an intrinsic excitation, as a function of the interaction 
strength. We can observe that in the case of two particles, both estimations, 
$\sqrt{M_{1}/M_{-1}}$ and $\sqrt{M_{3}/M_{1}}$, differ and neither have the 
same energy of the intrinsic one, indicating that there is more than one peak 
in the dynamic structure function. At $g=0$ , there is only one peak in which 
two states coexist, i.e. an intrinsic and a center-of-mass excitation, both 
estimates coincide and provide the exact excitation energy. Then the differences 
grow, indicating the existence of more than one peak in $S_F(E)$. In fact, 
there are higher intrinsic excitations whose strength decrease quickly besides 
the center-of-mass excitation which is fixed in the excitation energy 2. The 
behavior of this difference is a consequence of the reentrance phenomena 
mentioned above. For large $g$ the differences decrease again and for 
$g \rightarrow \infty $ both estimations coincide and provide the exact 
excitation energy which again is the same for the center of mass and the 
only excited intrinsic state, similarly to the non-interacting 
case Fig.~\ref{fig:SF_breathing2p}. When the number of particles increases, 
the strength of the center-of-mass peak decreases and the two ratios provide closer estimates which in turn are close 
to the main intrinsic excitation energy. This can be seen in Fig.~\ref{fig:SF_breathing4p}, 
where the intrinsic excitation is clearly dominant over the rest of 
excitations. Similar to the two particle case, the interaction strength region 
where the estimates are worse is where the intrinsic excitation energy 
takes its lower value.
 
\subsection{Interaction quench}

An alternative way to explore the internal excitations of an interacting 
many-body system is by performing an instantaneous quench of the interaction 
strength, which for  the two particle case has been studied
in~\cite{Schmelcher1,Sowinski2}, and for different confining potential in~\cite{Mistakidis1}. In current experiments
in ultracold atomic gases this is
feasible. Indeed, a great control on the interaction strength is achieved 
by means of Feshbach resonances~\cite{feshbach.reson.ultracold}.

A possible protocol is to prepare the system in its ground state for a certain 
value of the interaction strength, then the latter is suddenly changed and as a 
consequence the original state is non-stationary anymore and evolves in 
time, populating the new set of eigenstates corresponding to the new value 
of the interaction strength.

This way to probe the system has similarities with the breathing mode. It 
also preserves spin and parity. However, due to the invariance of the 
interaction under translations, there are no center-of-mass excitations.

\subsubsection{Time evolution of the perturbed system}

We want to study the time evolution of the ground state corresponding to a 
given value of $g$ after performing a sudden quench of the interaction 
strength to a new value $g_{\mathrm{new}}$. The initial ground state is 
not anymore the ground state of the new Hamiltonian. However we can 
express the old ground state in terms of the eigenvectors of the new 
Hamiltonian and  calculate its time evolution:
\begin{equation}\label{final_state_quench}
    \ket{\Psi(t=0)}=\sum_n c_n \ket*{\Psi^{'}_n}\,,
\end{equation}
where $\ket{\Psi(t=0)}$ is the ground state of the Hamiltonian before 
the quenching, and the states $\ket*{\Psi^{'}_n}$ are the eigenfunctions 
of the Hamiltonian after the quenching
\begin{equation}
    H^{'}\ket*{\Psi^{'}_n}=E^{'}_n\ket*{\Psi^{'}_n}\,,
\end{equation}
with eigenenergies $E^{'}_n$. The coefficients $c_n$ can be computed as:
\begin{equation}
    c_n=\bra*{\Psi^{'}_n}\ket*{\Psi(t=0)}\,.
\end{equation}
The time evolution of the state after the quenching is calculated by 
taking into account that the states $\ket*{\Psi^{'}_n}$ are the eigenstates 
of the new Hamiltonian, 
\begin{equation}
    \ket*{\Psi(t)}=\sum_n c_n e^{-i E^{'}_n t } \ket*{\Psi^{'}_n}\,.
\end{equation}
In turn, the eigenstates $\ket*{\psi^{'}_n}$ can be expressed in terms of 
the many-body basis used to describe our system:
\begin{equation}
    \ket*{\Psi^{'}_n}=\sum_i C_{n,i}\ket{\psi_i}\,,
\end{equation}
where $\ket{\psi_i}$ is a state of the many-body basis.

Knowing the time evolution  of the state we can study the time dependence of 
any observable. In particular we are going to consider the time evolution 
of the central density of our system:
\begin{equation}
    \rho (x=0,t)=\sum_{i,j}\sum_{k,l}\sum_{n,m}c_n^{*}c_m 
    e^{-i(E^{'}_m-E^{'}_n)t}C_{n,k}^{*}C_{m,l}
    \bra{\psi_k}a_j^{\dagger}a_i\ket{\psi_l}\Phi_i^{*}(0)\Phi_j(0)\,,
\end{equation}
where $\Phi_i(x)$ are the single-particle harmonic oscillator wave functions. 
Notice that the time evolution of the system is governed by $E^{'}_m-E^{'}_n$, 
involving all energy differences of the states of the new Hamiltonian that 
have non-zero overlap with the ground-state of the old Hamiltonian. Also important 
is to realize that the relevance of a given frequency $2\pi\nu_{n,m}=E_n-E_m$ 
depends on the product of the projection coefficients $c_n^{*}c_m$.

For instance, it is plausible to think that, for a small change of $g$, the original 
ground state should have a large overlap with the new ground state. In this situation the 
dominant frequencies will be associated to the differences $E^{'}_n-E^{'}_0$. In 
this sense, the frequencies would be similar to the frequencies involved in the 
calculation of the dynamical structure function associated to the breathing mode 
of $H^{'}$, with the exception that the excitations of the center of mass are not 
present.
\begin{figure}[t]
    \centering
    \includegraphics[width=\textwidth]{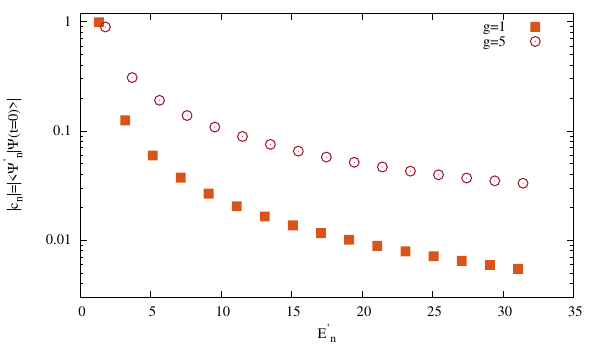}
    \caption{Projection of the initial ground state for  $g=0$ into the eigenstates 
    of $g=1$ and $g=5$ as a function of its energy, for a two particles systems with 
    zero total spin. The $y$ axis is in logarithmic scale.}
    \label{fig:projection_2p_0_1_5}
\end{figure}

In Fig.~\ref{fig:projection_2p_0_1_5}, we report, for the two-particle system, 
the projections of the ground state of the non-interacting system to the eigenstates 
of $H^{'}(g=1)$. 
We also report the projection for the quench $g=0\rightarrow 5$. 
In the case $g=1$ (small $g$) there 
is a dominant projection into the ground state of $H^{'}$ and then the value 
of the projections decreases rather quickly. For the case $g=5$, the largest 
projection is smaller than in the previous case but the projections to higher 
states decrease more slowly. Therefore for $g=5$ one also has relevant projections 
into higher excited states.

\subsubsection{Central density oscillations}

In this section we propose a method to determine the low energy spectrum of a 
trapped few-body system, by applying a sudden quench of the strength of the 
interatomic interactions. In principle, the procedure can be implemented experimentally.

One can prepare a system with $N$ particles trapped in a harmonic potential 
without interactions. Then apply to the system a sudden quench of the interaction, 
let the system evolve and measure the central density as a function of time. One 
should observe an oscillatory behavior. 
\begin{figure}[t]
    \centering
    \includegraphics[width=\textwidth]{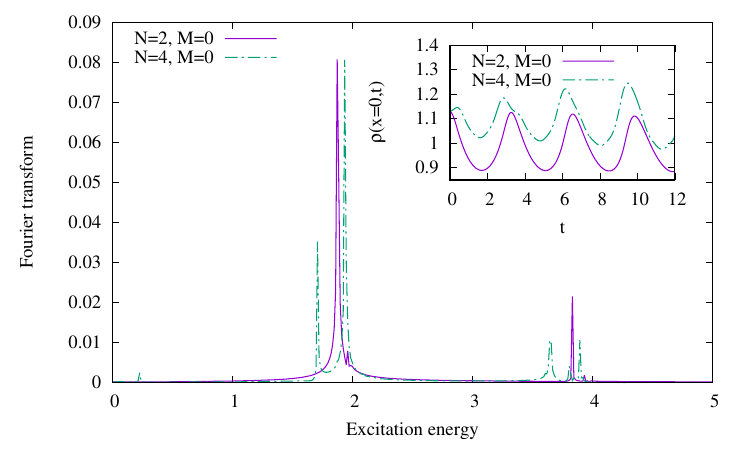}
    \caption{Frequency analysis of the time dependence of the central density after an interaction quench 
    from $g=0$ to $g=1$. The average density has been subtracted. We consider the cases $N=2$ (solid lines) and $N=4$ 
    (dot-dashed lines). In the inset we depict the first oscillations of the signal 
    used to compute the Fourier analysis shown in the main panels. For the Fourier analysis we used a time interval,  t=500. All magnitudes 
    in the figure are in harmonic oscillator units.}
    \label{fig:oscill_2p_0_1}
\end{figure}

In order to obtain information about the energy spectrum of the system at $g=1$ 
(the final value of the quench), we perform a Fourier analysis of the central 
density as a function of time, see Fig.~\ref{fig:oscill_2p_0_1}. In this way one 
obtains the characteristic oscillation frequencies which in turn are related 
to  differences of the energies of the system for $g=1$.  Therefore, from the 
frequencies of the Fourier analysis one can reconstruct the excitation energies 
$\Delta E = 2 \pi  \nu$, mainly with respect to the ground state. To extract 
the frequencies by Fourier analysis one needs to know the time evolution of 
the central density in an interval of time much longer than the one shown in 
the inset of Fig.~\ref{fig:oscill_2p_0_1}. Notice also that it is convenient 
to subtract the average value of the central density in order to avoid a peak 
with zero frequency in the Fourier analysis. Actually, this peak at $\nu=0$ 
does not provide any physical information about the excitation energy spectrum 
and could overlap with other peaks at low frequencies. We observe a dominant peak 
at low energy, and more peaks with smaller intensity at higher energies. Although 
the time interval used to perform the  Fourier analysis is very large, the 
Fourier frequencies still come out with an uncertainty width.

The location of the peaks allows us to recover the excitation energies of the 
low-excited states with respect to the ground state corresponding to $g=1$. 
However, to find the absolute energies of these states would be necessary to 
have access also to the ground-state energy of the system.

The figure reveals the existence of a dominant peak at an excitation energy 
a little smaller than 2, and other two main visible peaks at energies a little 
smaller than 4 and 6, with much smaller strength. All these peaks correspond 
to energy differences of the first excited states, which have non-zero overlap 
with the non-interacting ground state ($g=0$), and the ground state of the 
system corresponding to $g=1$, respectively. Notice that the reentrance 
phenomenon, discussed in the previous section, is reflected in the fact that 
the excitation energies are a little smaller than 2,4, and 6 respectively. It 
is also worth to mention the existence of a smaller but distinguishable peak, 
very close to the excitation energy 2, that corresponds to the difference 
between the second and first intrinsic excitations (see Fig.~\ref{fig:spectrum}). 
No center-of-mass excitations affect this analysis.

In the same figure, Fig.~\ref{fig:oscill_2p_0_1}, we present also the case of four 
particles and spin projection $M=0$. The shape of $\rho(x=0,t)$ shows also 
an oscillatory behavior, which should be determined by several frequencies. 
There are two dominant peaks which can be identified, by looking at the 
excitation spectrum of this system  for $g=1$ in Fig.~\ref{fig:4particlesspin}, 
with the lowest excitation energies with respect to the ground state. Actually, 
the small peak that appears at very low energy, corresponds to the energy 
difference of these two dominant peaks. Notice also that as the ground state 
has $S=0$, and the interaction is spin independent, the quench cannot connect 
states with different spins and therefore they do not participate in the 
time dependence of $\rho(x=0, t)$. As explained for the two-particle case, 
due to the translation invariance of the interaction, the  center-of-mass 
excitations do not play any role in the analysis. 
A similar procedure for the time dependence of the mean square 
radius of the system leads to a determination of the excitation energies 
consistent with the ones obtained by the analyzing the time dependence of 
the maximum of the density.

\clearpage
\section{Summary and conclusions}\label{chap6}

In this paper we have presented a quantum microscopic description of few 
spin $1/2$ fermions trapped in a 1D harmonic oscillator potential. The fermions interact 
through a contact interaction that, due to the Pauli principle, is active only 
between particles with opposite spin. Along the paper, we have mainly concentrated 
on repulsive interactions. These types of systems have already been realized 
experimentally and new experiments concerning the structure and mainly the dynamics 
are expected for the near future. Our main objective has been to study both static 
and dynamic properties of the system as a function of the interaction strength. 
This interaction, which provides a good modelization of  real interactions, 
can be experimentally controlled  by means of Feshbach resonances.

After a brief introduction, in the second  section, we have described the Hamiltonian 
of the system, that includes both the harmonic oscillator confining potential and 
the interaction between the fermions. Furthermore, two analytical limits have been 
discussed in detail: the non-interacting and the infinite interacting limits. In 
both cases, the many-body wave function, the energy and the density profile of 
the ground state have been derived. In addition, the second quantization formalism 
has been introduced as the framework for the calculation of the Hamiltonian.

In the third section we have described in detail the numerical procedures used 
in the paper, mainly the exact diagonalization techniques. Special attention has 
been devoted to the convergence problems associated with the dimension of the 
subspaces of the Hilbert space used to diagonalize the Hamiltonian. The 
benchmark of the analytical results for the two-particle system has been used 
to test the accuracy of the numerical methods.

In the next  section, we have studied several static properties of the ground 
state of the system, mainly the total energy and their contributions. The energy 
contributions have been used to check the fulfillment of the virial relation, 
obtaining very good results for repulsive interaction, specially in the 
low-interacting regime. The fulfilment of the virial relation reinforced 
the consistency of the calculations. We also have discussed  the density 
profile of the ground state. In the non-interacting and the infinite 
interacting cases our calculations are in perfect agreement with the analytical 
results presented in the first section. In the next step, we have computed 
the natural orbits and its eigenvalues. Their analysis reveals the presence 
of important correlations beyond mean field in the system when the interaction is turned on.

In the last section, we have studied several dynamical  properties. First we 
have studied the response of the system to  a mono-polar excitation over the 
ground state. We have computed the dynamic structure function associated to 
this perturbation, and using the knowledge of the energy spectra obtained in 
the previous section, we have identified the excited states. The  spin assignament
was useful to identify the excited states by the breathing mode. We have also 
calculated the energy momenta ($M_n$) associated to the dynamic structure 
function and  used the ratios $\sqrt{M_n/M_{n-2}}$ to estimate the average 
excitation energies. The differences between the ratios $\sqrt{M_n/M_{n-2}}$ 
for different $n$ indicates the presence  of more than one excited state in 
the dynamic structure function. In fact, for the two particle case there is 
more than one dominant peak in the response of the system. However, for 
larger number of particles, even if there are more excited states, the 
strength of the response is mainly concentrated in one dominant peak.

Finally, we have studied the effect of a quench of the interaction strength 
on the system. We have discussed the similarities and the differences between 
the quench and the breathing mode. In addition, we have proposed a possible 
experimental procedure to measure the excitation energies based on the 
measurement of the time evolution of the central density of the system after 
a quench of the interaction. Analyzing the frequencies of the Fourier transform 
of the time dependence of the oscillation of the central density after the 
quench one can obtain information about the excitation energies. We expect that 
the methods and results presented in the paper estimulate the use of exact 
diagonalization methods to study the dynamics and time evolution of systems 
with  a few fermions  in more complicated geometries and different interactions. 
\vspace{6pt}

\acknowledgments{We thank very useful correspondence from M. Simeon, T. Sowinski 
and N. Zinner. This work is partially funded by MINECO (Spain) Grants No. FIS2017-
87534-P and from the European Union Regional Development Fund within the
ERDF Operational Program of Catalunya (project QUASICAT/QuantumCat).}


%

\clearpage
\appendixtitles{no} 
\appendix

\section{Derivation of the virial theorem}\label{apendix1}

The energy of the system is given by the expectation values
$\expval{E}=\expval{T}+\expval{V_{\mathrm{ho}}}+\expval{V_{\mathrm{int}}}$, 
where the operators are defined as
\begin{equation}
    T=\sum_{i}-\frac{1}{2}\frac{\partial^2}{\partial x_i^2}\,,\quad
    \;V_\mathrm{ho}=\sum_i\frac{x_i^2}{2}\,,\quad V_\mathrm{int}
    =\sum_{j<i} g\delta(x_i-x_j)\,.
\end{equation}
We use the virial theorem to obtain a relation between the different energy 
contributions. The virial theorem is based on a scaling transformation of the 
many-body wave function and the transformation of the different energy contributions 
under the scaling transformations. We start by defining a wave function
\begin{equation}
    \Psi_\lambda(x_1,x_2,...,x_N)=\lambda^{N/2}\Psi(\lambda x_1,\lambda x_2,...,\lambda x_N)\,.
\end{equation}
This wave function has the same normalization than $\Psi$.

First, we compute the transformation of the kinetic energy,
\begin{equation}
    \expval{T(\lambda)}=\bra{\Psi_\lambda}\sum_i -\frac{1}{2}\frac{\partial^2}{\partial x_i^2}\ket{\Psi_\lambda} =\lambda^2\expval{T(\lambda=1)}\,.
\end{equation}
The harmonic oscillator energy under this transformation is
\begin{equation}
    \expval{V_{\mathrm{ho}}(\lambda)}=\bra{\Psi_\lambda}\sum_i \frac{x_i^2}{2}\ket{\Psi_\lambda}  = \lambda^{-2}\expval{V_{\mathrm{ho}}(\lambda=1)}\,.
\end{equation}
And finally, the interaction energy transforms as 
\begin{equation}
\begin{split}
   \expval{V_{\mathrm{int}}(\lambda)}&=\bra{\Psi_\lambda}\sum_{j<i} g\delta(x_i-x_j)\ket{\Psi_\lambda}\\
   &=\frac{N\left(N-1\right)}{2}g\lambda^N\int dx_1 ... dx_N \Psi(\lambda x_1,\lambda x_2,...,\lambda x_N)\delta\left(x_1-x_2\right)\Psi(\lambda x_1,\lambda x_2,...,\lambda x_N)\\
   &=\frac{N\left(N-1\right)}{2}g\lambda^N\int \frac{1}{\lambda^N}dy_1 ... dy_N \Psi( y_1, y_2,..., y_N)\lambda \delta\left(y_1-y_2\right)\Psi(y_1, y_2,...,y_N)\\
    &=\lambda\expval{V_{\mathrm{int}}(\lambda=1)}\,,
\end{split}
\end{equation}
where we have implemented the change of variable $\lambda x_i = y_i$ for all $i$.
The expectation value of the Hamiltonian with the scaled wave function can be written as
\begin{equation}
    \expval{E_\lambda}=\bra{\Psi_\lambda}H\ket{\Psi_\lambda}
    =\lambda^2\expval{T(\lambda=1)}+\lambda^{-2}\expval{V_{\mathrm{ho}}}+\lambda\expval{V_{\mathrm{int}}}\,,
\end{equation}
and taking into account that at $\lambda=1$ the energy has a stationary point, i.e.,
\begin{equation}
    \left.\frac{\partial \expval{E_\lambda}}{\partial \lambda}\right|_{\lambda=1}=0\,,
\end{equation}
we derive the virial relation
\begin{equation}
\begin{split}
    \left.\frac{\partial \expval{E_\lambda}}{\partial \lambda}\right|_{\lambda=1}=0=
    &\left.2\lambda\expval{T(\lambda=1)}+-2\lambda^{-3}
    \expval{V_{\mathrm{ho}}(\lambda=1}+\expval
    {V_{\mathrm{int}}(\lambda=1)}\right|_{\lambda=1}\\
    =&2\expval{T}-2\expval{V_{\mathrm{ho}}}+\expval{V_{\mathrm{int}}}=0\,.
    \end{split}
\end{equation}

\section{Evaluation of the one-body matrix elements}\label{apendix2}

The Hamiltonian in second quantization has been written in the harmonic 
oscillator basis and requires the knowledge of the single-particle matrix 
elements of both the harmonic trapping potential and the kinetic 
energy. Just for completeness, we include here a brief derivation.

The states of the harmonic oscillator single-particle basis read 
 \begin{equation}
     \ket{n}=\frac{1}{\sqrt{2^n n!\sqrt{\pi}}}H_n (x) e^{-x^2/2}
     \ket{\chi_{m_n}}=A_n H_n (x) e^{-x^2/2}\ket{\chi_{m_n}}\,,
 \end{equation}
 where $A_n$ is a normalization constant, which depends on $n$. On the 
 other hand $\ket{\chi_{m}}$ is a single-particle spin-$1/2$ wave function, 
 with spin projection $m$.
 
 In order to operate with the Hermite polynomials, we use the following 
 properties:
 \begin{itemize}
     \item[a)] The $m$-th derivative of a Hermite polynomial,
 \begin{equation}\label{hermite_m_derivate}
     H_n^{(m)}=2^m \frac{m!}{(n-m)!}H_{n-m}(x)\,,
 \end{equation}
 \item[b)] The recurrence relation,
\begin{equation}\label{recurrence_hermite}
     H_{n+1}(x)=2xH_n(x)-2nH_{n-1}(x)\,,
 \end{equation}
 \item[c)] The orthogonality of the Hermite polynomials
 \begin{equation}\label{orthogonality}
    \int_{-\infty}^\infty H_m(x) H_n(x) e^{-x^2}=\delta_{m,n}2^n n! \sqrt{\pi} \,,
\end{equation}
\item[d)] The $n$-th power of $x$ expressed in terms of Hermite polynomials
 \begin{equation}\label{powerhermite}
    x^n=\frac{n!}{2^n}\sum_{m=0}^{n/2} \frac{1}{m!(n-2m)!}H_{n-2m}\,,
\end{equation}
\item[e)] The product of two Hermite polynomials as a function of 
the sum of Hermite polynomials
\begin{equation}\label{2hermite_sumhermite}
    H_m(x)H_n(x)=2^nn!\sum_{r=0}^n\frac{m!}{(n-r)!(m-n+r)!}
    \frac{H_{m-n+2r}(x)}{2^r r!} , n\leq m \,.
\end{equation}
\end{itemize}

First, we derive the expression of the one-body harmonic oscillator matrix elements, that reads,
\begin{equation}\label{expvalvho0}
    \bra{j}V_{\mathrm{ho}}\ket{i}=\bra{j}\frac{x^2}{2}\ket{i}\,.
\end{equation}
Using Eq.~(\ref{powerhermite}) we can write $x^2$ as
\begin{equation}
    x^2=\frac{1}{2}\left(\frac{1}{2}H_2+H_0\right)\,.
\end{equation}
Then, writing explicitly the Eq.~(\ref{expvalvho0}), and taking into account Eq.~(\ref{2hermite_sumhermite}) and the orthogonality Eq.~(\ref{orthogonality}) we get,
\begin{equation}
\begin{split}
    \bra{j}V_{\mathrm{ho}}\ket{i}=&\bra{j}\frac{1}{4}
    \left(\frac{1}{2}H_2+H_0\right)A_i H_i e^{-x^2/2}\ket{\chi_{m_i}}\\
    =&\bra{j}\frac{A_i}{4}\left(2n(i-1)H_{i-2}+(2i+1) H_i
    +\frac{1}{2}H_{i+2}\right)e^{-x^2/2}\ket{\chi_{m_i}}\\
    =&\frac{A_i A_j}{4}\int dx H_j(x)\left(2n(i-1)H_{i-2}(x)+(2i+1) H_i(x)
    +\frac{1}{2}H_{i+2}(x)\right)e^{-x^2}\bra{\chi_{m_i}}\ket{\chi_{m_j}}\,.
\end{split}
\end{equation}
Finally, the harmonic oscillator matrix element  is given by 
\begin{equation}\label{harmonicoscillatorenergy}
    \bra{j}V_{\mathrm{ho}}\ket{i}=\frac{1}{4}\left(\sqrt{i(i-1)}
    \delta_{j,i-2}+(2i+1)\delta_{j,i}+\sqrt{(i+2)(i+1)}\delta_{j,i+2}\right)\delta_{m_i,m_j}\,.
\end{equation}

Next, we derive the expression of the one-body kinetic matrix elements, that read,
\begin{equation}
    \bra{j}T\ket{i}=\bra{j}\left(-\frac{1}{2}\frac{\partial^2}{\partial x^2}\right) \ket{i}\,.
\end{equation}
Taking the first and second derivatives of the harmonic oscillator wave function 
and using Eq.~(\ref{hermite_m_derivate})
 \begin{equation}\label{1derhermite}
     \frac{\partial}{\partial x}H_n(x)= 2nH_{n-1}(x)\,,
\end{equation}
\begin{equation}\label{2derhermite}
     \frac{\partial^2}{\partial x^2}H_n(x)= 4n(n-1)H_{n-2}(x)\,,
\end{equation}
we can express the kinetic energy as
 \begin{equation}
     \bra{j}T\ket{i}=\bra{j}\frac{A_i}{2}e^{-x^2/2}
     \left(-4i(i-1)H_{i-2}+4x\,i\,H_{i-1}+H_i-x^2\,H_i\right)\ket{\chi_{m_i}}\,.
 \end{equation}
Using Eq.~(\ref{recurrence_hermite}) we obtain:
\begin{equation}
   \bra{j}T\ket{i}=\bra{j}A_ie^{-x^2/2}\left(i+\frac{1}{2}-\frac{x^2}{2}\right)H_i(x)\ket{\chi_{m_i}}\,,
\end{equation}
and taking the results of the harmonic oscillator potential energy Eq.~(\ref{harmonicoscillatorenergy}), the kinetic energy matrix elements are expressed as
\begin{equation}
    \bra{j}T\ket{i}=\frac{1}{4}\left(-\sqrt{i(i-1)}\delta_{j,i-2}+(2i+1)\delta_{j,i}-\sqrt{(i+2)(i+1)}\delta_{j,i+2}\right)\delta_{m_i,m_j}\,.
\end{equation}
Obviously, both the kinetic energy and the harmonic oscillator potential are diagonal in the spin projection.

\clearpage
\reftitle{References}



\sampleavailability{Samples of the compounds ...... are available from the authors.}


\end{document}